\documentclass[twocolumn]{aastex63}
\usepackage{amsmath}

\newcommand{\diff}{d}

\newcommand{\rfb}{r_{\rm fb}}
\newcommand{\Nsc}{N_{\rm sc}}
\newcommand{\Lbox}{L_{\rm box}}
\newcommand{\delx}{\Delta x}
\newcommand{\xsci}{\mathbf{x}_{{\rm sc},i}}
\newcommand{\mdot}{\dot{M}_w}
\newcommand{\pdot}{\dot{p}_w}
\newcommand{\Lwind}{\mathcal{L}_w}
\newcommand{\mcloud}{M_{\rm cloud}}
\newcommand{\rcloud}{R_{\rm cloud}}
\newcommand{\REC}{R_{\rm EC}}

\newcommand{\nhat}{\mathbf{\hat{n}}}
\newcommand{\rhat}{\mathbf{\hat{r}}}
\newcommand{\Vinj}{V_{\rm inj}}
\newcommand{\foldedness}{\left\langle\nhat \cdot \rhat\right\rangle}

\newcommand{\reff}{\mathcal{R}_{b}}
\newcommand{\rfree}{\mathcal{R}_{\rm f}}
\newcommand{\dotreff}{\dot{\mathcal{R}}_{b}}
\newcommand{\Vbub}{V_{b}}
\newcommand{\Ebub}{E_{b}}
\newcommand{\Abub}{A_{b}}
\newcommand{\pr}{p_{r}}
\newcommand{\ms}{M_{*}}
\newcommand{\sfe}{\varepsilon_{*}}
\newcommand{\rhobar}{\bar{\rho}}
\newcommand{\ecool}{\dot{E}_{\rm cool}}
\newcommand{\Ebkgnd}{\dot{E}_{\rm bkgnd}}
\newcommand{\ratr}{\mathcal{S}}

\newcommand{\lcool}{\ell_{\rm cool}}

\newcommand{\tcool}{t_{\rm cool}}
\newcommand{\fcool}{f_{\rm cool}}

\newcommand{\Esh}{E_{\rm sh}}
\newcommand{\Ersh}{E_{\rm r, sh}}
\newcommand{\Etsh}{E_{\rm turb, sh}}
\newcommand{\fwind}{f_{\rm wind}}
\newcommand{\strfunc}{\left\langle\mathbf{v}_t^2 (\ell) \right\rangle^{1/2}}
\newcommand{\vw}{{\cal{V}}_w}

\newcommand{\vt}{v_{ t}}

\newcommand{\pc}{\mathrm{\, pc}}
\newcommand{\yr}{\mathrm{\, yr}}
\newcommand{\Myr}{\mathrm{\, Myr}}
\newcommand{\kms}{\mathrm{\, km\, s^{-1}}}

\received{XXX}
\revised{XXX}
\accepted{XXX}

\submitjournal{ApJ}

\shorttitle{Efficiently Cooled Wind Bubbles: Simulations}
\shortauthors{Lancaster et al.}

\begin{document}

\title{Efficiently Cooled Stellar Wind Bubbles in Turbulent Clouds II.\\ Validation of Theory with Hydrodynamic Simulations}

\correspondingauthor{Lachlan Lancaster}
\email{lachlanl@princeton.edu}

\author[0000-0002-0041-4356]{Lachlan Lancaster}
\affiliation{Department of Astrophysical Sciences, Princeton University, 4 Ivy Lane, Princeton, NJ 08544, USA}

\author[0000-0002-0509-9113]{Eve C. Ostriker}
\affiliation{Department of Astrophysical Sciences, Princeton University, 4 Ivy Lane, Princeton, NJ 08544, USA}

\author[0000-0001-6228-8634]{Jeong-Gyu Kim}
\affiliation{Department of Astrophysical Sciences, Princeton University, 4 Ivy Lane, Princeton, NJ 08544, USA}

\author[0000-0003-2896-3725]{Chang-Goo Kim}
\affiliation{Department of Astrophysical Sciences, Princeton University, 4 Ivy Lane, Princeton, NJ 08544, USA}

\begin{abstract}

In a companion paper, we develop a theory for the evolution of stellar wind 
driven bubbles in dense, turbulent clouds.  This theory proposes that turbulent 
mixing at a fractal bubble-shell interface leads to highly efficient cooling, in 
which the vast majority of the input wind energy is radiated away.  This energy 
loss renders the majority of the bubble evolution momentum-driven rather than 
energy-driven, with expansion velocities and pressures orders of magnitude lower 
than in the classical \citet{Weaver77} solution.  In this paper, we validate our 
theory with three-dimensional, hydrodynamic simulations.  We show that extreme 
cooling is not only possible, but is generic to star formation in turbulent 
clouds over more than three orders of magnitude in density.  We quantify the few 
free parameters in our theory, and show that the momentum exceeds the wind input 
rate by only a factor $\alpha_p \sim 1.2-4$.  We verify that the bubble/cloud 
interface is a fractal with dimension $\sim 2.5-2.7$. The measured turbulent 
amplitude ($v_t \sim 200-400 \kms$) in the hot gas near the interface is shown 
to be consistent with theoretical requirements for turbulent diffusion to 
efficiently mix and radiate away most of the wind energy.  The fraction of energy 
remaining after cooling is only $1-\Theta \sim 0.1-0.01$, decreasing with time, 
explaining observations that indicate low hot-gas content and weak dynamical 
effects of stellar winds.
\end{abstract}


\section{Introduction}
\label{sec:intro}

Feedback from massive stars is thought to be the dominant process setting 
the lifetime efficiency of star formation on the scale of individual molecular 
clouds \citep{McKeeOstriker2007,LadaLada03}. Luminous hot stars use their 
strong radiation to disperse the surrounding dense gas in several ways: 
imparting primarily-outward photon momentum to gas and dust where stellar 
UV/optical radiation is first absorbed (direct radiation pressure); imparting 
preferentially-outward photon momentum to dust where diffuse infrared is 
absorbed (reprocessed radiation pressure); ionizing and heating gas, which 
leads to over-pressured expansion of ionized gas and creates a rocket effect 
on neutral structures where photoevaporation occurs (pressure of photo-ionized 
gas); or directly depositing momentum in the star's own atmosphere, resulting 
in high-velocity stellar winds which shock and transfer momentum to the ambient 
gas. It is much debated in the literature which of these mechanisms dominates 
in which situations within the star-forming interstellar medium (ISM) 
\citep{KMBH19}. While several recent efforts have used numerical radiation 
hydrodynamic simulations for in-depth studies of the effects of radiation 
feedback on star-forming clouds \citep{Dale12,Dale13a,Walch12,Raskutti16,raskutti17,Howard17,JGK18,JGK19,JGK20,Grudic20,Geen20a,Geen20,Fukushima20}, there have 
been fewer detailed numerical studies of stellar wind feedback (reviewed 
below).  In this paper, we focus on the thermal and dynamical effects of 
stellar winds driven by star clusters in their natal molecular clouds.

In an accompanying paper \citep[][hereafter, Paper I]{PaperI} we have 
theoretically investigated the dynamics of expanding stellar wind bubbles in the 
case that most of the wind's energy is radiated away.  There, we argue that 
the interface between hot bubbles and the surrounding gas is subject to 
strong turbulent mixing, which subsequently leads to efficient radiative 
cooling in intermediate-temperature gas. We also argue that the surface of 
the hot bubble is a fractal, and the large associated area enhances cooling. 
We hypothesize that efficient cooling leads to the dominant phase of bubble 
evolution being momentum-driven rather than energy-driven. In this 
work, we support the theoretical model of Paper I with a suite of 
three-dimensional (3D) hydrodynamic simulations, which we analyze to provide 
quantitative corroboration of the arguments made there.  Each simulation 
follows the expansion of a hot bubble driven into the surrounding dense, 
turbulent ISM by a constant luminosity point source.

Several numerical studies have previously explored the dynamical effects of 
main sequence stellar winds on their environment. These have included 
one-dimensional (1D) numerical models 
\citep{GarciaSegura96,GarciaSegura96a,SilichTT13,Krause16,Fierlinger16,Rahner17} 
some of which have incorporated phenomenological accounts of three-dimensional 
processes such as energy leakage \citep{HCM09} and turbulent diffusivity that 
leads to cooling \citep{ElBadry19}. These problems have also been explored in 
two-dimensional (2D) studies, which have sought to examine filamentation at 
reduced computational expense \citep{Wunsch08,Ntormousi11,Dwarkadas13}. Finally, 
numerical investigations of stellar wind feedback effects in 3D have ranged from 
targeted studies of winds 
\citep{RogersPittard13, Wareing17,Krause13,KrauseDiehl14,DaleBonnell08,Dale13a},
to comparisons of winds and photo-ionization heating over a range of parameter 
space \citep{Dale14,Geen15a,Haid18,Geen20}. These simulations have contributed 
to assessments of the relative contribution of winds in a number of scenarios 
and over a range of densities and cloud masses. Some have additionally tried to 
constrain the relative contribution of leakage and turbulent mixing.

In this paper, we explore a set of questions related to the detailed structure 
and evolution of bubbles driven by stellar winds in turbulent clouds.  Our goals 
are to characterize the dependence of bubble evolution on input wind power and 
ambient cloud properties, and to explain the physical mechanisms controlling the 
evolution. Using numerical simulations, we quantify 
(1) the temporal evolution of wind bubble sizes and the energy and momentum that 
bubbles/shells contain; (2) the properties of the turbulence that drives mixing 
at the interface between a bubble's hot/diffuse interior and cool/dense shell; 
(3) the fractal structure and foldedness of the cooling interface; and (4) the 
total energy losses to cooling due to turbulent mixing. Confirming the theory 
laid out in Paper I, we shall show that for the conditions within star-forming 
clouds, the vast majority of wind energy is expected to be lost due to turbulent 
mixing followed by radiative cooling, and that the large area associated with 
the fractal geometry of the bubble/cloud interface is crucial to enhancing 
losses. Our physical picture of turbulent, cooling mixing layers is much 
informed by insights from recent numerical and analytic investigations of 
the Kelvin-Helmholtz instability in the context of multiphase galactic winds 
and the circumgalactic medium \citep{GronkeOh18,FieldingFractal20,Tan20}.

The numerical model we adopt is intentionally idealized. It is designed to 
provide a testing ground for the theory presented in Paper I that represents 
key real-world features (to the extent that is computationally practical), 
while still having that theory be applicable. To that end, we (i) have ignored 
the effects of star formation that is extended in both space and time, (ii) 
approximated the winds as having constant mechanical luminosity, (iii) ignored 
the effects of magnetic fields and stellar radiation, and (iv) adopted the 
assumption of collisional and ionization equilibrium for computing cooling 
losses in ionized gas, and a simple analytic fit for cooling in warm/cold 
neutral gas. Thus, our simulations are designed to test our theoretical 
predictions, rather than faithfully represent the real world. By tackling 
an idealized numerical problem, however, we are able to validate our 
theoretical framework and quantitative predictions that can be extrapolated 
beyond the necessarily limited scope of any single simulation suite.

The structure of this paper is as follows. In \autoref{sec:theory} we briefly 
review our theory for evolution of wind-driven bubbles. The reader is referred 
to Paper I for a complete description. In \autoref{sec:methods} we describe 
our numerical methods for implementing winds in the \textit{Athena} code, as 
well as our simulation set-up and the parameters of our model suite.  We 
present the analysis of our results in \autoref{sec:results} and conclude with 
a summary of our findings in \autoref{sec:conclusion}.

\section{Review of Paper I Theory: Wind Bubble Dynamics with Efficient Cooling}
\label{sec:theory}

Our theory is developed in full in 
Paper I; for convenience we briefly summarize key features here. Our global
dynamical evolution model is based on the premise that energy losses in the 
bubble/shell interface are so great that the bubble is effectively momentum-driven 
rather than effectively energy-driven\footnote{
We use the term ``momentum-driven'' (``energy-driven'') to denote a solution in 
which momentum (energy) increases linearly in time.  The 
\citet{Castor75}/\citet{Weaver77} solution is also internally energy-conserving 
in the sense that there are no radiative energy losses from the hot bubble or its 
interface with the shell; however, the leading shock is assumed to be fully radiative 
so that a fraction $27/77=0.35$ of the input wind energy is radiated away as the 
shocked and accelerated ambient gas cools to join the exterior of the shell.}. We 
term this global solution the ``Efficiently Cooled (EC)" stellar wind-driven bubble.

In the EC model, there is a point source of constant mechanical luminosity 
$\Lwind$ and mass loss rate $\mdot$, which injects into the surrounding 
cloud a wind with velocity
\begin{equation}
    \vw \equiv \left({\frac{2\Lwind}{\mdot}}\right)^{1/2} \, 
\end{equation}
and momentum input rate 
\begin{equation}\label{eq:pdot}
    \pdot \equiv \vw\mdot = 2 \Lwind/\vw = \left({2\Lwind\mdot}\right)^{1/2}  \, .
\end{equation}
The momentum of the system is mostly contained in the bubble shell, and increases 
linearly in time as
\begin{equation}
    \label{eq:pEC}
    \pr = \alpha_p p_{\rm EC} \equiv \alpha_p \pdot t 
    =2\alpha_p \frac{\Lwind}{\vw}t \, .
\end{equation}
This allows for amplification by a factor $\alpha_p$ (assumed order-unity in the 
EC theory) above the value $p_{\rm EC}$ that corresponds to the momentum originally 
injected in the wind, which would apply if there is no thermal energy buildup within 
the bubble (maximally efficient mixing/cooling).

We define $\reff$ as the radius of the sphere that has the same volume as 
the bubble:
\begin{equation}
    \label{eq:reff_def}
    \reff \equiv \left(\frac{3\Vbub}{4\pi} \right)^{1/3} \, .
\end{equation}
Under the assumptions that the background density variations are statistically 
isotropic and homogeneous, and that the bubble's momentum is given by 
\autoref{eq:pEC}, the bubble's effective 
radius evolves in time as
\begin{equation}
    \label{eq:rbub_ecw}
    \reff (t)  = 
    \left(\frac{3\alpha_R \alpha_p}{2\pi} 
    \frac{\pdot t^2}{\rhobar} \right)^{1/4} 
    \equiv (\alpha_R\alpha_p)^{1/4} \REC \, .
\end{equation}
Here, $\alpha_R$ is another order-unity parameter that accounts for 
geometry; $\REC$ denotes the solution for an exactly spherical bubble 
expanding in a uniform ambient medium with momentum increasing at a 
rate $\dot p_w$. The scaling $\reff \propto t^{1/2}$ is shallower than 
$\reff \propto t^{3/5}$ for the classical energy-driven solution. 

The total energy contained within the bubble interior is
\begin{equation}
    \label{eq:ebub_relation}
    \Ebub = \frac{1}{2}\pdot \ratr \reff  \, ,
\end{equation}
where $\ratr\ge 1$ quantifies the enhancement in energy above the case where 
the bubble is fully occupied by the free wind. For a spherically expanding 
bubble, we can write $\ratr$ in terms of $\alpha_p$ (see Equations A13 and 
A16 of Paper I), with $\ratr\approx\alpha_p$ within 6\%. 

The predicted pressure in the post-shock gas of the wind is
\begin{eqnarray}
    \label{eq:pressure}
    P_b &=& \frac{3}{16\pi} \frac{\pdot}{\rfree^2}\nonumber \\
    &=& \frac{3}{16\pi} \frac{\pdot}{\reff^2} 
    \left\{\frac{2}{3}\alpha_p + \left[ \left(\frac{2}{3}\alpha_p\right)^2 - \frac{1}{3}\right]^{1/2}\right\}
\end{eqnarray}
where $\rfree$ is the radius of the free wind region, defined 
analogously to \autoref{eq:reff_def}.  From Equation A14 of Paper I, the term in curly braces can be approximated as $(\reff/\rfree)^2\approx (3\alpha_p-1)/2$  within 4\% for $1<\alpha_p<4$.

The radial kinetic energy in the bubble's shell is  given by
\begin{equation}\label{eq:EKr}
    \Ersh = \frac{\alpha_p }{4\alpha_R}\pdot\reff .
\end{equation}
The shell will also have turbulent motion, such that 
$\Esh=\Ersh+ \Etsh$ is its total kinetic energy. We describe the 
level of turbulent energy relative to the radial kinetic energy as
\begin{equation}
    \label{eq:fturb_def}
    f_{\rm turb} \equiv \frac{\Etsh}{\Ersh} \, .
\end{equation}
Paper I shows that the above formulae can be combined to obtain  a prediction 
for the fraction of input energy retained after cooling,
\begin{equation}
    \label{eq:Theta_cool_prediction}
    1 - \Theta = 
    \left(\frac{1}{2}(1 + f_{\rm turb})\frac{\alpha_p}{\alpha_R} +\ratr \right)
    \frac{\dotreff}{\vw} \, ,
\end{equation}
where $\Theta\equiv \dot{E}_{\rm cool}/\Lwind$ is the complementary 
fraction lost to cooling.  To the extent that the terms in parentheses 
in \autoref{eq:Theta_cool_prediction} are relatively constant, we would 
expect $1 -\Theta \propto t^{-1/2}$.

In Paper I, we hypothesize that the bubble-shell interface is a fractal. 
Under this assumption, the bubble's total area when measured on scale 
$\ell$ becomes 
\begin{equation}
    \label{eq:frac_area}
    \Abub(\reff;\ell) \equiv 4\pi \alpha_A \reff^2 
    \left(\frac{\reff}{\ell} \right)^d \, ,
\end{equation}
where $d$ is known as the ``excess fractal dimension'' of the surface 
and $\alpha_A$ is an order-unity parameter meant to account for 
any minor inconsistencies with this model.

Paper I argued that instabilities at the interface between the shocked wind 
and the surroundings drive turbulence that feeds off the kinetic energy 
in the shocked wind.  The turbulence in the hot gas is assumed to follow a 
power-law form, such that $\vt(\ell) = \vt(L)(\ell/L)^p$.  Accounting for both 
the turbulent velocity and fractal area of the interface, the effective enthalpy 
flux at scale $\ell$ is then $\Phi_{\rm cool} \sim (5/2) P v_{\rm equiv}$, where  
\begin{equation}\label{eq:veff}
    v_{\rm equiv} (\ell)= \vt(\ell) (\reff/\ell)^d\, 
\end{equation} 
defines an equivalent radial velocity through the turbulent mixing/cooling 
boundary layer for a bubble of size $\reff$.  This enthalpy flux represents 
the capacity for mixing and radiative cooling, increasing to smaller scale 
until reaching the scale where $\vt(\ell_{\rm cool})\tcool =\ell_{\rm cool}$, 
where 
\begin{equation}\label{eq:tcool}
    \tcool \equiv \frac{P}{n^2 \Lambda(T_\mathrm{pk} )} 
    =  \frac{(k_B T_\mathrm{pk})^2}{P\Lambda(T_\mathrm{pk})} \, ,
\end{equation}
for $T_\mathrm{pk}$ the temperature of peak cooling.  With pressure in the 
shocked gas given by \autoref{eq:pressure}, $\tcool \propto \reff^2$.

Finally, we note that Paper I outlined the conditions under which the EC regime 
should apply. One estimate, from comparing to the solution from \citet{ElBadry19}, 
gives an upper limit on the energy retention fraction for the EC solution to be 
valid: 
\begin{equation}
    \label{eq:cooling_condition}
    1 - \Theta < 4\left(\frac{5}{6}\right)^{1/4}
    \frac{\dot{R}_{\rm EC}}{\vw} \, .
\end{equation}
Another limit is that the velocity of shocked hot gas flowing into the boundary 
layer, 
\begin{equation}\label{eq:vrb}
v_{\rm hot}(\reff)=v_{\rm ps}\left(\frac{\rfree}{\reff}\right)^2\sim \frac{\vw}{6\alpha_p -2},
\end{equation}
cannot exceed the equivalent velocity that turbulent diffusion can accommodate, 
given in \autoref{eq:veff}. 

\section{Numerical Methods and Models}
\label{sec:methods}

To test the theory presented in Paper I, 
we use the \textit{Athena} code \citep{Stone08_Athena} in running a series of 
three-dimensional (3D) hydrodynamic simulations of constant luminosity 
stellar winds injected into turbulent clouds. We do not include any magnetic 
fields in the present simulations, but we do include cooling in the gas that 
is implemented as part of the \textit{Athena}-TIGRESS code base for the 
star-forming multiphase ISM \citep{CGK_TIGRESS1,CGK_TIGRESS2}. This implements 
cooling in an operator-split manner following \citet{KoyamaInutsuka02} at 
$T<10^{4.2}{\rm K}$ and \citet{SutherlandDopita93} at $T>10^{4.2}{\rm K}$. 

There is also uniform background heating of 
$2\times 10^{-26} \, {\rm erg}\,{\rm s}^{-1}\,{\rm H}^{-1}$ which decreases 
in hot, ionized gas; more details can be found in Section 2.3.1 of 
\citet{CGK_TIGRESS1}. All of our simulations are run using the linearized 
Roe Riemann solver \citep{Roe81}, second-order spatial reconstruction, and 
the unsplit van Leer integrator \citep{StoneGardiner09}.

\subsection{Wind Injection}
\label{subsec:windmethod}

Here we describe our treatment of stellar winds, which we implement in 
{\it Athena} in an operator-split fashion. 
We provide a quantitative test of this implementation in \autoref{app:validation}.
Each simulation contains a single wind source, which we refer to as a 
``star particle,'' although in practice it represents a stellar cluster.
This source particle is put in ``by hand'' and 
does not exert any gravitational force on the surrounding gas.
Our approach employs a hybrid thermal/kinetic 
energy injection scheme which interpolates between pure thermal injection close 
to a source and pure kinetic injection towards the edge of the feedback region. 
There are multiple benefits to this hybrid energy injection approach. From a 
physical perspective, it more directly represents the reality of a bubble
driven by a cluster of massive stars,
wherein the majority of the energy close to the center of the bubble is 
thermal 
(due to shock thermalization through colliding winds of individual stars) while it is mostly kinetic towards the edges (where this thermal energy has managed to drive expansion).
From a numerical perspective, for a 
purely kinetic feedback implementation the vector momentum field cannot be 
properly resolved immediately adjacent to the source. By transitioning to 
thermal energy injection near the source, it is only necessary to resolve a 
scalar field. An advantage of the hybrid approach over purely thermal energy 
injection is that the latter can require a larger spatial scale for pressure 
gradients to accelerate the flow modeling the primary wind.

Given a star particle at position $\mathbf{x}_{\rm star}$ within the domain 
of the simulation, we deposit energy in the  grid cells surrounding the star 
particle using a subcell method based on the implementation of \citet{Ressler20}.  
We specify $\rfb$, the radius of the spherical feedback region, and $\Nsc$ is
the number of subcells per grid cell along a grid-aligned direction. When 
initializing the simulation we loop through a cube of $N_{\rm sc,tot}^3$ 
subcells where 
\begin{equation}
    N_{\rm sc,tot} = 2\Nsc\left\lceil\rfb/\delx \right\rceil
\end{equation}
is the total number of subcells that could possibly lie within the 
feedback sphere along a grid-aligned direction and $\lceil\cdot \rceil$
denotes the ceiling function. The subcells within this cube are equally 
spaced along each grid-aligned direction between 
$-\delx \lceil \rfb/\delx\rceil$ and $+\delx \lceil \rfb/\delx\rceil$.

For each subcell in this cube with position $\xsci$ relative to the source,
we determine if $|\xsci| < \rfb +\delx/2\Nsc$ 
(i.e. if the subcell is within half a subcell spacing of the feedback 
radius 
or fully within the feedback radius). If it is, we then record the position of the subcell 
$\xsci$ for use within the full simulation. In order 
to account for the subcell volume only partially overlapping the actual 
feedback region, we additionally assign each subcell a weight 
$w_{{\rm sc},i}$ according to
\begin{equation}
    w_{{\rm sc},i} = \left\{\begin{array}{ll}
        1 & \quad |\xsci| \leq \rfb - \frac{\delx}{2\Nsc} \\
        \frac{1}{2} \left(\frac{2\Nsc(\rfb - |\xsci|)}{\delx} + 1\right) & 
        \quad |\xsci| > \rfb - \frac{\delx}{2\Nsc}
    \end{array}{}\right. \, .
\end{equation}
This formula is simply a linear interpolation between 1 and 0 over the 
radial range of $\pm$ half a subcell spacing from the feedback radius, which 
approximates the true volume fraction \citep{CubeSphereIntersect}. We 
additionally store the total effective volume that wind energy is injected 
into, defined as
\begin{equation}
    \Vinj = \left( \frac{\delx}{\Nsc}\right)^3 
    \sum_i w_{{\rm sc},i} \, .
\end{equation}

Once the positions and weights of the subcell template are initialized, 
they are used for the hybrid thermal/kinetic energy injection. 

Given a mechanical luminosity $\Lwind$ and mass loss rate $\dot{M}_w$ 
for the wind, the mean mass density $\rho_w$ and energy density 
$\epsilon_w$ to be injected in a time $\Delta t$ are 
\begin{equation}
    \rho_w = \frac{\dot{M}_w \Delta t}{\Vinj} \, \, , \, \, 
    \epsilon_w = \frac{\Lwind \Delta t}{\Vinj} \, .
\end{equation}
For a given subcell in our template, indexed by $i$, we determine the 
grid cell in which it resides at location $\mathbf{x}_{\rm star} + \xsci$, 
and increment the mass density by 
$\delta \rho_i = w_{{\rm sc}, i} \rho_w/\Nsc^3$. For the purposes of 
the current paper, $\mathbf{x}_{\rm star}$ is always at the origin.

The total energy density in the target cell is similarly incremented by
$\delta \epsilon_i = w_{{\rm sc},i} \epsilon_w / \Nsc^3$.  To compute the 
injected momentum, we specify the  variable $f_{\rm kin}$ at a given 
subcell position $\xsci$ according to
\begin{equation}
    \label{eq:fkin_def}
    f_{\rm kin} (|\xsci|) = {\rm max}\left(1, |\xsci|/\rfb \right) \, ,
\end{equation}
Once $f_{\rm kin}$ is determined we define the 
injected momentum density as
\begin{equation}
    \delta \mathbf{q}_i = 
    \left( 2f_{\rm kin}(|\xsci|)\delta \epsilon_i\delta \rho_i \right)^{1/2}
    \frac{\xsci}{|\xsci|} \, .
\end{equation}

As was noted in \citet{WallMacLow20}, the presence of momentum and mass within 
the grid cell to which we are adding these quantities means that the kinetic energy 
added by the above change in momentum density \textit{is not} 
$f_{\rm kin}\delta \epsilon_i$, but actually somewhat less depending on what 
the initial density and momentum of the cell are. 
The discrepancy between the actual injected kinetic energy density and 
$f_{\rm kin}\delta \epsilon_i$ is especially important at the early times 
of wind onset, when the surrounding density is high, but becomes 
negligible once the wind has evacuated the feedback region. 
In the case where there is negligible pre-existing energy and momentum,
our prescription deposits  75\% of the total energy  as kinetic energy.

In order to make sure that our results are not affected by the details of our 
feedback mechanism, we run  tests with $f_{\rm kin}=0$ (purely thermal feedback) 
in Appendix~\ref{app:thermal_feedback}. This scenario might be closer to the truth 
of a cluster of stars whose colliding winds shock and turn initially kinetic 
energy into bulk thermal energy. We found that this pure thermal feedback 
scenario simply leads to a minor delay in the wind's evolution, but broadly 
the results remain unchanged.

When we inject the wind mass into the feedback region we add an equal amount 
of ``mass'' to a passive scalar variable, which is simply passively advected with the 
gas.  We use this scalar to track the mixing of the wind material with the 
surrounding gas as well as a means of separating mass that has been swept up 
by the wind from the ambient medium, as we will explain in \autoref{sec:results}.

\begin{deluxetable*}{cccccc}
\tablecaption{Parameters of simulation suite.\label{tab:sim_params}}
\tablewidth{0pt}
\tablehead{
\colhead{Cloud Mass} & Cloud Radius & \colhead{$\overline{n}_{\rm H}$}  & $\vt$ &
$\Delta x$\tablenotemark{a} & Resolution\tablenotemark{a}\\
\colhead{$[M_{\odot}]$} & $[{\rm pc}]$ & \colhead{$[{\rm cm}^{-3}]$} & 
$[{\rm km}\, {\rm s}^{-1}]$ & $[{\rm pc}]$ & }
\startdata
$5\times 10^4$ & 20  & 43.1  & 3.59 & $0.15$ & $256^3$\\
$5\times 10^4$ & 10  & 345.  & 5.08 & $0.08$ & $256^3$\\
$5\times 10^4$ & 5   & 2760  & 7.18 & $0.04$ & $256^3$\\
$5\times 10^4$ & 2.5 & 22800 & 10.2 & $0.02$ & $256^3$\\
$10^5$ & 20  & 86.3  & 5.08 & $0.08$ & $512^3$\\
$10^5$ & 10  & 690.  & 7.18 & $0.04$ & $512^3$\\
$10^5$ & 5   & 5520  & 10.2 & $0.02$ & $512^3$\\
$10^5$ & 2.5 & 44200 & 14.4 & $0.01$ & $512^3$\\
$5\times 10^5$ & 20  & 431  & 11.4 & $0.15$ & $256^3$\\
$5\times 10^5$ & 10  & 3450  & 16.1 & $0.08$ & $256^3$\\
$5\times 10^5$ & 5   & 27600  & 22.7 & $0.04$ & $256^3$\\
$5\times 10^5$ & 2.5 & 228000 & 32.1 & $0.02$ & $256^3$
\enddata
\tablecomments{Each model is run with three different values for the mass 
of the wind source particle, $M_*/\mcloud\equiv \sfe = 0.01,0.1,$ and $1$.
\tablenotetext{a}{Only highest resolution is provided.}}
\end{deluxetable*}

\subsection{Model Parameters and Set-Up}
\label{sec:run_describe}

Our goal is to span a large range of the potential parameter space for 
stellar winds originating from young star clusters within massive 
molecular clouds, which is where most star formation takes place.  To that 
end, we consider three cloud masses $\mcloud$ of $5\times 10^4 M_{\odot}$, 
$10^5 M_{\odot}$, and $5\times 10^5 M_{\odot}$ and four different cloud 
radii $\rcloud$ of $20 \, {\rm pc}$, $10 \, {\rm pc}$, $5 \, {\rm pc}$, and 
$2.5\, {\rm pc}$.  The largest radii represent ``normal'' star-forming GMCs, 
cases with intermediate radii may represent infrared dark clouds (IRDCs), 
and cases with the smallest radii may represent extremely dense 
cluster-forming clumps within larger GMCs or the natal clouds in which super 
star clusters are born in starbursting galactic centers.

Each choice of $\mcloud$ and $\rcloud$ corresponds to a choice of 
mean mass density $\rhobar$ as
\begin{equation}
    \rhobar = \frac{3\mcloud}{4\pi \rcloud^3}  = \mu m_p 
    \overline{n}_{\rm H}\; ,
\end{equation}
where $\mu_{\rm H}=1.4271$ is the mean molecular weight of the gas, $m_p$ is 
the mass of a proton, and $\overline{n}_{\rm H}$ is the mean number density 
of the Hydrogen nuclei. Rather than creating spherical clouds within our 
cubic domain, we apply uniform density $\rhobar$ everywhere within a domain 
of side length $\Lbox = 2\rcloud$.  The ``global cloud'' parameters are 
just used for reference in setting the ambient conditions with which the 
wind interacts.  

The grid has uniform spatial resolution in each dimension with 
$\Lbox/\delx =128,\, 256, \, 512$ for the simulations with 
$\mcloud=10^5\, M_{\odot}$ and only $\Lbox/\delx = 128,\, 256$ for the simulations at 
higher and lower cloud mass. The parameters describing the simulation initial 
density conditions and numerical resolution are outlined in 
\autoref{tab:sim_params}. Only the highest resolution is listed for each model. 

We initialize all the simulations at a  uniform pressure of 
$P/k_B = 3\times 10^4 \,{\rm cm}^{-3} \, {\rm K}$. Since we evolve the 
simulations long enough for thermal relaxation to occur before initiating the 
wind, this particular choice of initial pressure is not important to our 
results. We also ran test simulations at a initial pressure of 
$P/k_B = 3\times 10^3 \,{\rm cm}^{-3} \, {\rm K}$ and found no differences. 

Each simulation is initialized with a turbulent velocity field which is a 
Gaussian random field with power spectrum $|{v}_k|^2 \propto k^{-4}$ for
$2 \le k L_{\rm box}/(2\pi) \le 64$. For higher resolution runs, the 
realization of the power spectrum uses an identical set of modes to the 
lower-resolution case. The amplitude of the turbulence is chosen so that 
the initial kinetic energy per unit mass is equivalent to twice that of 
the gravitational potential for a sphere of radius $\rcloud$ and 
mass $\mcloud$, i.e. 
\begin{equation}
    \tilde{E}_{\rm kin,i} = 2|\tilde{W}_{\rm sphere}| = \frac{6G\mcloud}{5\rcloud}\, ,
\end{equation}
where the tildes are used to indicate energy per unit mass.
We note, however, that self-gravity is not included in the simulations.

For each simulation, we allow the turbulent velocity field to evolve and 
decay, without any additional driving, until the average kinetic energy per 
unit mass has decayed to $\tilde{W}_{\rm sphere}$ (i.e. it has been reduced by half). 
At the end of this initial evolution each simulation has a turbulent velocity
scale $\vt=(\tilde{E}_{\rm kin,i})^{1/2}$. For each model, $\vt$ is listed in 
\autoref{tab:sim_params}.

At the same time as turbulence is decaying, the Reynolds stresses create 
inhomogeneous density structure throughout the simulation domain.
The turbulent decay period is generally one tenth of the initial flow 
crossing time ($\Lbox/v_{t,i}$) of the simulation domain, amounting to 
0.03-0.7 Myr.  Over this period, the thermal energy also relaxes, with 
temperature approaching thermal equilibrium; in the cold gas this is 
$\sim 30-200 \, {\rm K}$.

After the initial period of turbulent decay, we create a star particle at the 
center of the simulation domain of mass $\ms = \sfe\mcloud$. The parameter 
$\sfe$ can be thought of as akin to a total star formation efficiency. For 
each set of initial conditions laid out in \autoref{tab:sim_params} we
run simulations at $\sfe = 0.01,0.1,$ and $1$.  The star 
particles inject energy and mass proportional to their own mass 
in accordance with the method described in \autoref{subsec:windmethod}, 
with a feedback radius $\rfb = \Lbox/40$ at all resolutions, meaning 
that the feedback radius is resolved by a minimum of 3.2 resolution 
elements at the lowest resolution. We test this choice of feedback radius 
in Appendix~\ref{app:rfb_test} and find it does not affect our results.

\begin{figure}
    \centering
    \includegraphics{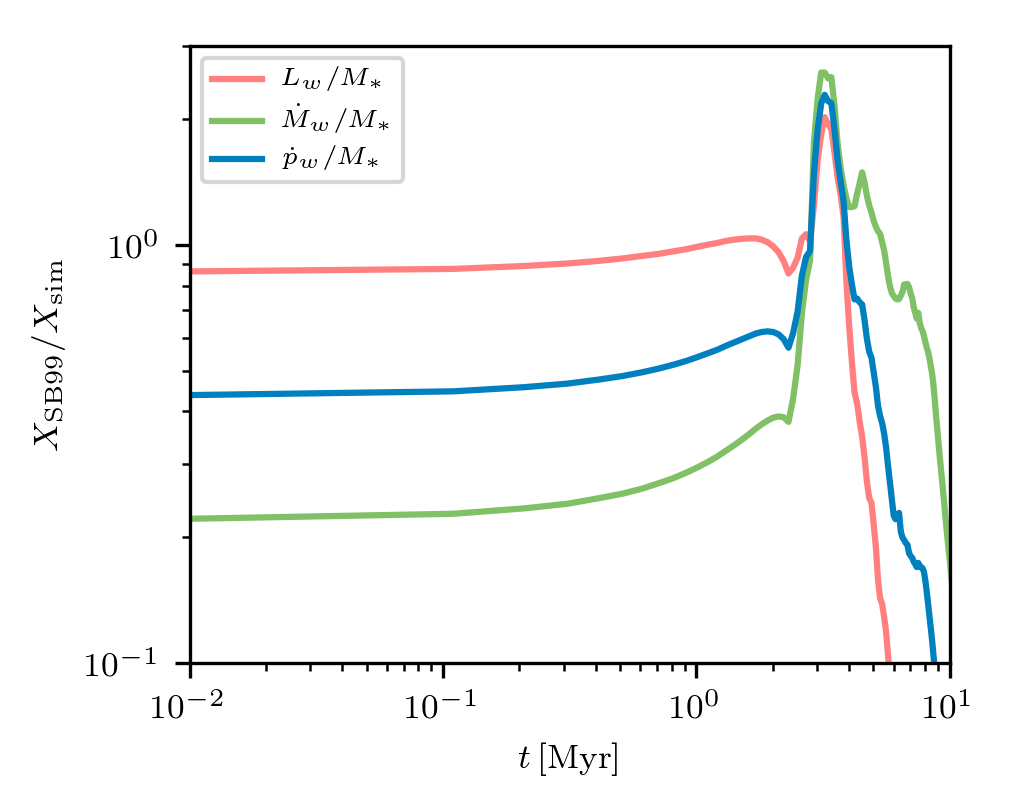}
    \caption{Ratio of the wind parameters derived using the population 
    synthesis code STARBURST99 (SB99) to the constant values for each 
    parameter that we adopt in our simulations.The SB99 mass-loss rate 
    and momentum input rate are respectively a factor $\sim 5$ and 
    $\sim 2$ lower than the ones we adopt.  We validate that using the 
    SB99 values does not change our conclusions 
    (see Appendix~\ref{app:reduced_mass_loss}).}
    \label{fig:sb99_comp}
\end{figure}

We adopt a constant wind luminosity per unit mass of 
$\Lwind/\ms = 10^{34}\, {\rm erg} \, {\rm s}^{-1} \, {M_{\odot}^{-1}}$ and 
mass loss rate per unit mass $\dot{M}_w/\ms = 10^{-2} \, {\rm Myr}^{-1}$.
This mass loss rate is about a factor of five larger than that expected 
for a standard solar metallicity population, with a Kroupa IMF 
\citep{KroupaIMF} as calculated by STARBURST99 (SB99) \citep{SB99}.  A 
comparison between the constant values that we adopt and those determined 
from SB99 is given in \autoref{fig:sb99_comp}.  We use the larger mass 
loss rate because this reduces the temperature of the shocked wind, yielding 
a less stringent time step.

Our adopted wind parameters  correspond to a wind velocity of 
$\vw = (2 \Lwind/\dot{M}_w)^{1/2}=1780 \, {\rm km} \, {\rm s}^{-1}$ 
and a momentum injection rate per unit mass of 
$\dot{p}_w/\ms = \vw \dot{M}_w/M_*=17.8 \,{\rm km} \,{\rm s}^{-1} \,{\rm Myr}^{-1}$. 
At the lower mass loss rates consistent with the SB99 calculations, 
the wind velocity (momentum injection rate) is larger (smaller) by a factor
$\sim \sqrt{5}\approx 2.2$, i.e. 
$\vw\sim 3750 \kms$ and $\dot{p}_w/\ms \sim  8.6 \kms \, \Myr^{-1}$.
This only corresponds to a difference in radial evolution 
(according to our theory) of a factor of $5^{1/8}\approx 1.2$, so 
this should not effect the interpretation of our results. We have 
additionally run simulations with a standard mass loss rate and have 
obtained consistent results, as described in Appendix~\ref{app:reduced_mass_loss}.

\begin{figure*}
    \centering
    \includegraphics[width=\textwidth]{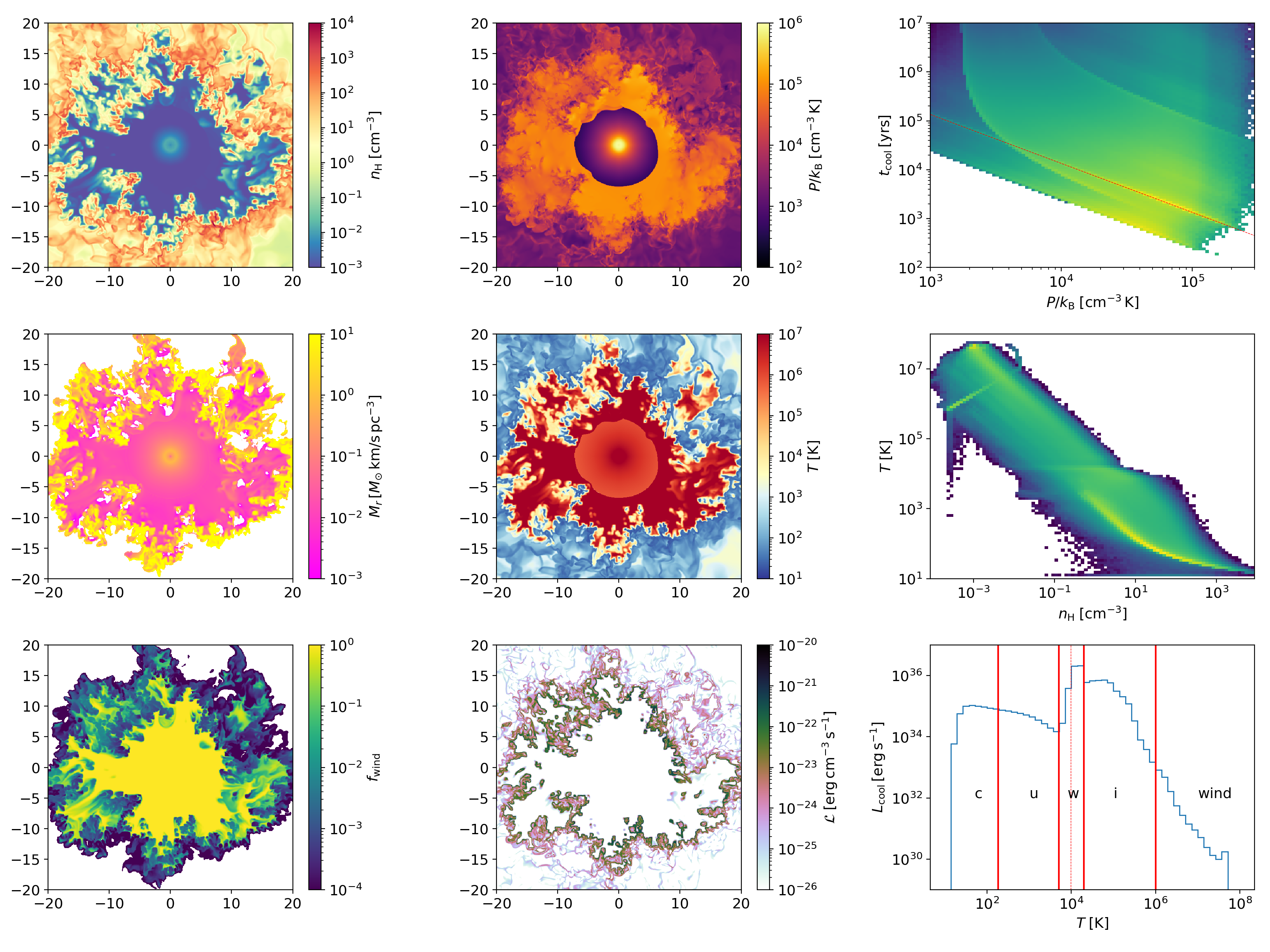}
    \caption{An example snapshot of the wind bubble structure and phase 
    properties at time $t=2.23 \, {\rm Myr}$ after the initiation of the wind
    for the model with $\mcloud = 10^5\, M_{\odot}$, $\rcloud = 20\, {\rm pc}$, 
    $\sfe = 1\%$, and $\Lbox/\delx = 512$ simulation. The left and center 
    columns show slices through the $z=0$ plane, while the right column 
    displays distributions. From left to right and top to bottom, panels show: 
    gas number density, pressure, a cooling-weighted phase diagram of cooling 
    time vs. pressure, radial momentum density, temperature, a volume-weighted 
    temperature-density phase diagram, the wind mass fraction ($\fwind$), the 
    cooling rate per unit volume, and a cooling-weighted histogram of the gas 
    temperature distribution (which depicts the total cooling occurring in 
    each temperature bin).  In the top-right panel, the highest concentration 
    of cooling is in gas that has $\tcool \propto P^{-1}$; this slope is 
    indicated by a dashed red diagonal line for the coefficient in 
    \autoref{eq:tcool} corresponding to $T_{\rm pk} = 9650 \, {\rm K}$. The 
    last panel uses vertical lines to delineate the phases described in 
    \autoref{tab:phase_defs}, where the subscripts from the table are indicated. 
    The label ``wind'' includes both the free (f) and post-shock (ps) wind phases.  
    In this panel, $T_{\rm pk}$ is marked as a vertical dashed line within the 
    ``w'' phase.
    }
    \label{fig:example_slice_R20}
\end{figure*}

\begin{figure*}
    \centering
    \includegraphics[width=\textwidth]{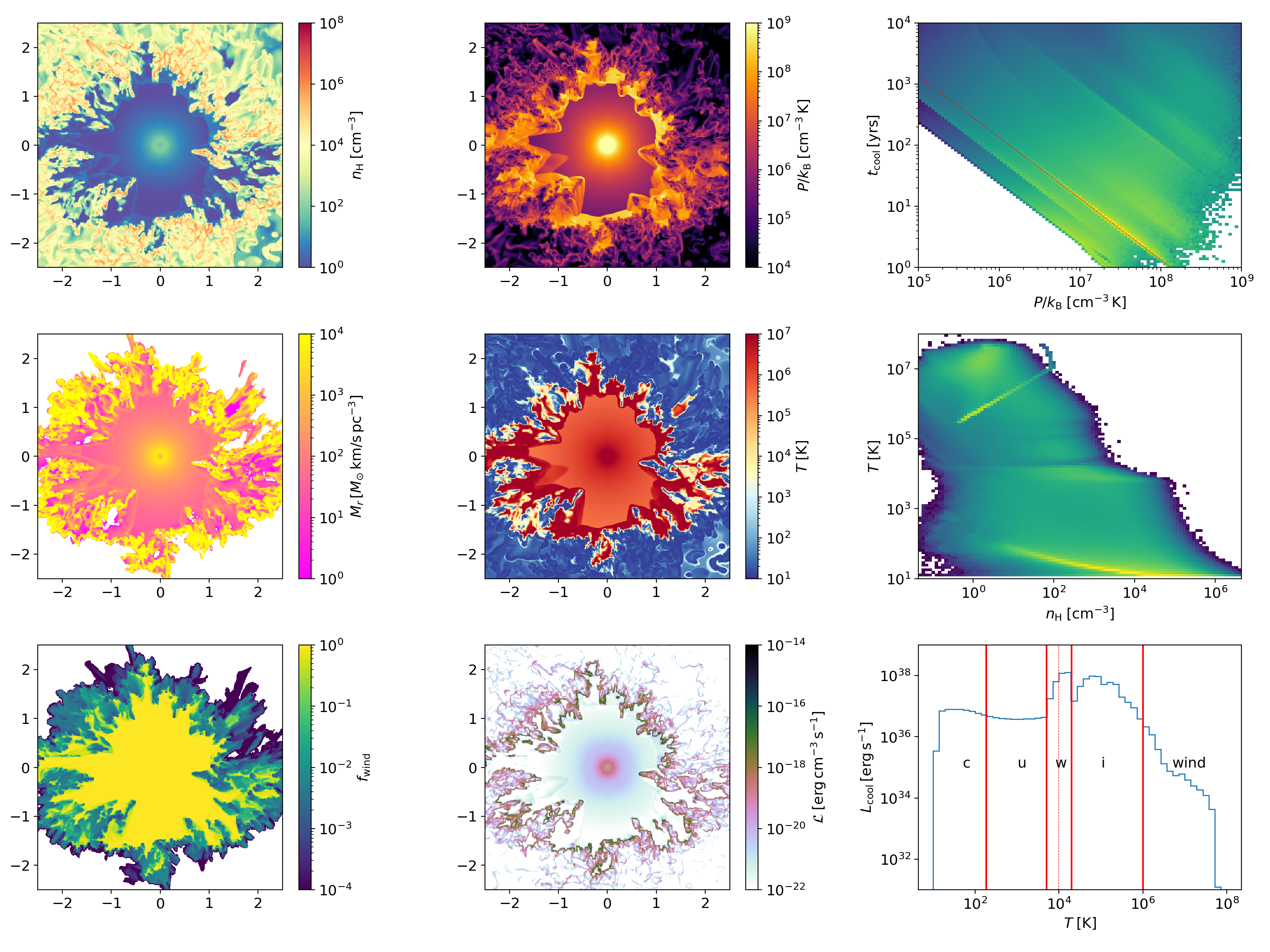}
    \caption{Same as \autoref{fig:example_slice_R20} but for the model with 
    $\mcloud = 10^5\, M_{\odot}$, $\rcloud = 2.5\, {\rm pc}$, $\sfe = 100\%$, 
    and $\Lbox/\delx = 512$ simulation at $t=0.125\, {\rm Myr}$. All quantities 
    are as specified in \autoref{fig:example_slice_R20}, though the range of 
    many quantities are quite different.}
    \label{fig:example_slice_R2p5}
\end{figure*}

\section{Numerical Results}
\label{sec:results}

Here we lay out the results of the simulations described in 
\autoref{sec:run_describe}.  In our presentation, we shall 
focus on cases with $\mcloud = 10^5\, M_{\odot}$ and use the 
remaining simulations for validation of the generality of our 
results, as summarized in Appendix~\ref{app:different_mcloud}.

\subsection{Cloud and Bubble Structure and Thermal Distributions}
\label{sec:structure}

Example snapshots from our simulations are provided in 
\autoref{fig:example_slice_R20} ($R = 20\, {\rm pc}$ and 
$\sfe = 1\%$ model) and \autoref{fig:example_slice_R2p5} 
($R = 2.5\, {\rm pc}$ and $\sfe = 100\%$ model) in order to 
provide a more concrete reference. These two models are 
deliberately chosen to span the range of cloud size and 
star formation efficiency parameters. 

In both figures, the ``wind-blown bubble'' is the low-density, 
high-temperature region filling the central region of the 
simulation domain, with fingers extending outward. Outside of the 
bubble, the inhomogeneous density structure generated by the background 
turbulence in the cold cloud is evident;  higher density portions of 
the ambient gas are also left behind in the bubble interior, although 
they are ablated over time by KH instabilities as the high-velocity 
wind flows past them.   

As expected, the cooling rate (bottom center panel) is greatest at 
the (fractal) interface between the hot bubble and cool shell, where 
mixing is driven by turbulence. The outward momentum originally carried 
by the wind is deposited at the bubble boundary by interface mixing, 
where it builds up in the expanding cool shell (middle left panel). The 
degree of mixing is also evident in the fraction of wind gas in each 
cell (bottom left panel).

\autoref{fig:example_slice_R20} and \autoref{fig:example_slice_R2p5} 
also include (right column) information on the statistical distribution 
functions of gas in density,  temperature, pressure, and cooling time. 
The top right panel shows that the highest concentration of cooling is 
in gas with   $\tcool \propto P^{-1}$, as predicted in Paper I (see 
definition in \autoref{eq:tcool}) .  Quantitatively, we find that 
\begin{equation}\label{eq:tcool_sim}
    \tcool = 135 \yr  
    \left(\frac{P/k_B}{10^6 \, {\rm K} {\rm cm}^{-3}}\right)^{-1}
\end{equation}
follows well the prominent linear feature at short cooling time for 
both cases, as well as our other models. The corresponding $T_{\rm pk}$ 
would be $9650 \, {\rm K}$. 

\begin{deluxetable*}{ccccc}
    \tablecaption{Definitions of gas phases.\label{tab:phase_defs}}
    \tablewidth{0pt}
    \tablehead{
    \colhead{Phase} &  \colhead{Temperature Condition} 
    &  \colhead{Velocity Condition}& \colhead{Subscript}& \colhead{Schematic Region}}
    \startdata
    Free Hyper-Sonic Wind & $T>10^5\, {\rm K}$ & $v_r > \vw /2$  & ${\rm f}$ & ${\rm(i)}$\\
    Shocked Stellar Wind & $T>10^6\, {\rm K}$ & $v_r < \vw /2$  & ${\rm ps}$ & ${\rm(ii)}$\\
    Ionized Gas & $2\times 10^4\, {\rm K}<T<10^6\, {\rm K}$ & $v_r < \vw /2$   
    & ${\rm i}$  & ${\rm(iii)}$\\
    Warm Neutral Gas & $5050\, {\rm K}<T<2\times 10^4\, {\rm K}$ & & ${\rm w}$ & ${\rm(iii)/(iv)}$\\
    Thermally Unstable Gas & $184\, {\rm K}<T<5050\, {\rm K}$ &  & ${\rm u}$& ${\rm(iii)/(iv)}$\\
    Cold Neutral Gas & $T<184\, {\rm K}$ &  & ${\rm c}$& ${\rm (iv)}$\\
    \enddata
    \tablecomments{Names assigned to gas phases based on temperature and 
    velocity conditions, with (second-from-right column) subscript used to denote 
    each phase. The final column indicates the rough correspondence 
    between gas phases and locations in the schematic Figure 1b of Paper I. }
\end{deluxetable*}

\subsection{Gas Phase Definitions}
\label{subsec:phase_def}

We define several different gas phases as an aid in quantifying the 
structure and evolution of the wind-driven bubbles. These definitions 
can be thought of as an expansion on the phases laid out for the 
classical stellar wind bubble in Paper I, in 
order to better account for the cooling of the gas. These definitions 
(see \autoref{tab:phase_defs}) are based on the temperature $T$ and 
radial velocity $v_r$ of the gas. The table also lists subscripts used 
to denote quantities associated with each phase (such as volume, 
momentum, energy, etc.). These subscripts label the thermal phases of 
the wind in the bottom right panels of \autoref{fig:example_slice_R20} 
and \autoref{fig:example_slice_R2p5}.

The first two phases listed in \autoref{tab:phase_defs} are analogous 
to the two wind phases described in \citet{Weaver77}. The Ionized 
Gas is contained within the cooling layer, and cooling in this phase 
is dominated by collisional excitation of transitions of H, He, C, N, 
and O (primarily). This phase is produced via mixing and subsequent 
cooling of shocked wind and shell gas, and would not exist as part of 
the undisturbed/ambient ISM in either the uniform or turbulent cases.

The Warm Neutral Gas has cooling dominated by collisionally-excited 
Ly$\alpha$ emission and recombination on dust grains.  The Thermally 
Unstable Gas is in the range between the stable equilibrium warm and cold 
phases in the static case, for our adopted cooling function.  This phase is 
continually populated because of the turbulence in the system.  Finally, at 
the lowest temperature there is the Cold Neutral Gas, which in reality would 
include both atomic and molecular phases, but here we do not follow the 
detailed chemistry.

The final three phases would all usually exist to varying degrees as 
part of the background in a turbulent, dense cloud. We wish to separate 
the portions that are ISM gas that has been shocked and then cooled after 
being swept into the expanding bubble shell from the portions that are undisturbed 
ambient gas. We do this using cuts based on 
the fraction of the mass in a given cell that originates in the wind, 
$\fwind$. The wind mass is tracked using a scalar that is injected with 
the wind (see \autoref{subsec:windmethod}) and then passively advected. 

For each of the Ionized, Warm, Thermally Unstable, and Cold phases 
of the gas we track all quantities of interest in gas with 
$\fwind > 10^{-3}, 10^{-4}, $ and $ 10^{-5}$ (each subsequent 
selection being a superset of the previous selection). Unless 
otherwise stated, we use $\fwind>10^{-4}$ to separate the swept-up 
and ambient parts of these phases, as this gives the best agreement 
between different methods of measuring the total cooling, as described 
in \autoref{subsec:energetics}. The tracking of gas properties at  
different $\fwind$ values also allows us to quantify the mixing of 
the wind with the turbulent gas.

As is evident in the bottom right panels of \autoref{fig:example_slice_R20} 
and \autoref{fig:example_slice_R2p5}, the majority of the cooling 
is occurring in the Warm and Ionized phases. These are the phases that 
primarily occupy the boundary region between the bubble and surrounding 
cloud, as is clear from the temperature and cooling slices in these 
figures.

\subsection{Bubble Evolution Comparisons}
\label{subsec:bubble_evol}

Here we present results from our simulations of the temporal evolution of 
the radial momentum of the gas ($\pr$), the wind bubble's effective radius 
($\reff$),  and the interior bubble energy ($\Ebub$).  We compare these 
results with the theory developed in Paper I and reviewed in \autoref{sec:theory}.  
When measuring quantities in the shell of the wind bubble, we include the 
measurement of this quantity in all gas with a wind mass fraction greater
than $10^{-4}$. We found this was the best way of differentiating between 
the swept-up shell and the background gas.

In this section we show results for the $\mcloud = 10^5\, M_{\odot}$ cases, 
while results for other cloud masses are presented in 
Appendix~\ref{app:different_mcloud} (\autoref{fig:m5e4_summary} and 
\autoref{fig:m5e5_summary}).

\begin{figure*}
    \centering
    \includegraphics{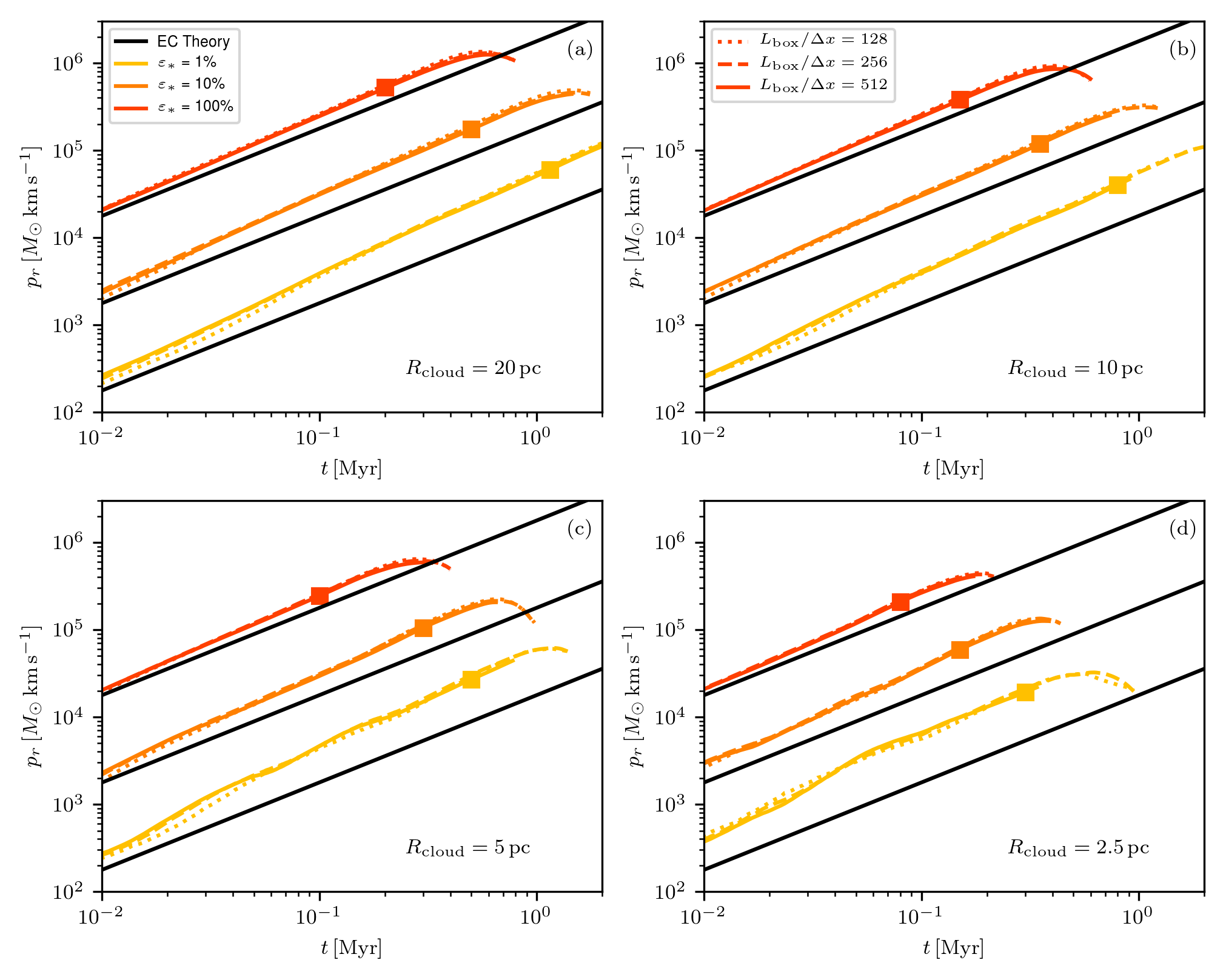}
    \caption{Evolution of the total radial momentum of the bubble (interior 
    plus shell) in each of our simulations for cases with cloud 
    mass $\mcloud = 10^5 M_{\odot}$. Model cloud radii (from least dense, 
    $\rcloud = 20\, {\rm pc}$, to most dense, $\rcloud = 2.5\, {\rm pc}$) 
    are given within the individual panels (a) to (d); see 
    \autoref{tab:sim_params} for parameters.   For each case we show 
    results for star cluster particles of mass 
    $\ms = 10^3 M_{\odot}, \, 10^4 M_{\odot}, \,$and$\, 10^5 M_{\odot}$ 
    (corresponding to star formation efficiencies of 
    $\sfe = 1\%,\,10\%, \,$and$\, 100\%$) in yellow,  orange, and 
    red respectively. At each value of $\rcloud$ and $\sfe$ we show three 
    separate lines (solid, dashed, dotted) corresponding to varying 
    resolution, as shown in the key.  For each simulation with $512^3$ 
    cells, we mark the time at which the first wind-polluted gas exits 
    the simulation domain $t_{\rm esc}$ with a square, and only show results 
    from each simulation up until $4 t_{\rm esc}$ (though not all of the 
    highest resolution simulations were run for this long).  For each value of 
    $\rcloud$ and $\sfe$ we show $\pdot t$ (which should apply when cooling 
    is maximally efficient, per \autoref{eq:pEC}) in black.}
    \label{fig:pr_comp}
\end{figure*}

\subsubsection{Shell Momentum}
\label{subsubsec:shell_momentum}
As is evident in \autoref{fig:example_slice_R20} and 
\autoref{fig:example_slice_R2p5}, most of the radial momentum is 
carried by the dense shell of swept-up gas, thus we define the 
momentum as:
\begin{equation}
    \label{eq:sim_pr_def}
    \pr \equiv p_{r, {\rm i}} + p_{r, {\rm w}}
    + p_{r, {\rm u}} + p_{r, {\rm c}}.
\end{equation}
As mentioned above, only cells with a wind mass fraction greater than 
$10^{-4}$ are included in this definition. We found no significant changes 
in our measurement when considering $\fwind>10^{-5}$ gas or even $\fwind>0$ 
gas, 
which we took to indicate that there was no significant momentum 
carried in ``completely unpolluted'' gas.
In \autoref{fig:pr_comp} (for the $\mcloud = 10^5\, M_{\odot}$ models), 
we compare this measurement of the momentum in the simulations with 
$\pr = \pdot t$, as predicted by the EC theory with $\alpha_p=1$ 
(\autoref{eq:pEC}).

The results at different resolutions show that the radial momentum 
carried by the bubble is extremely well converged in our simulations. 
Overall, results are within a factor of 1.2-4 of the EC prediction. The 
results are quantitatively closest to $\pdot t$ for more luminous winds 
(corresponding to higher star formation efficiency). 

The momentum in excess of $\pdot t$ can be attributed to nonzero 
buildup of thermal energy in the shocked wind that aids in driving 
expansion (see Appendix of Paper I). This is parameterized in 
our theory by $\alpha_p$ in \autoref{eq:pEC}.

\begin{figure*}
    \centering
    \includegraphics{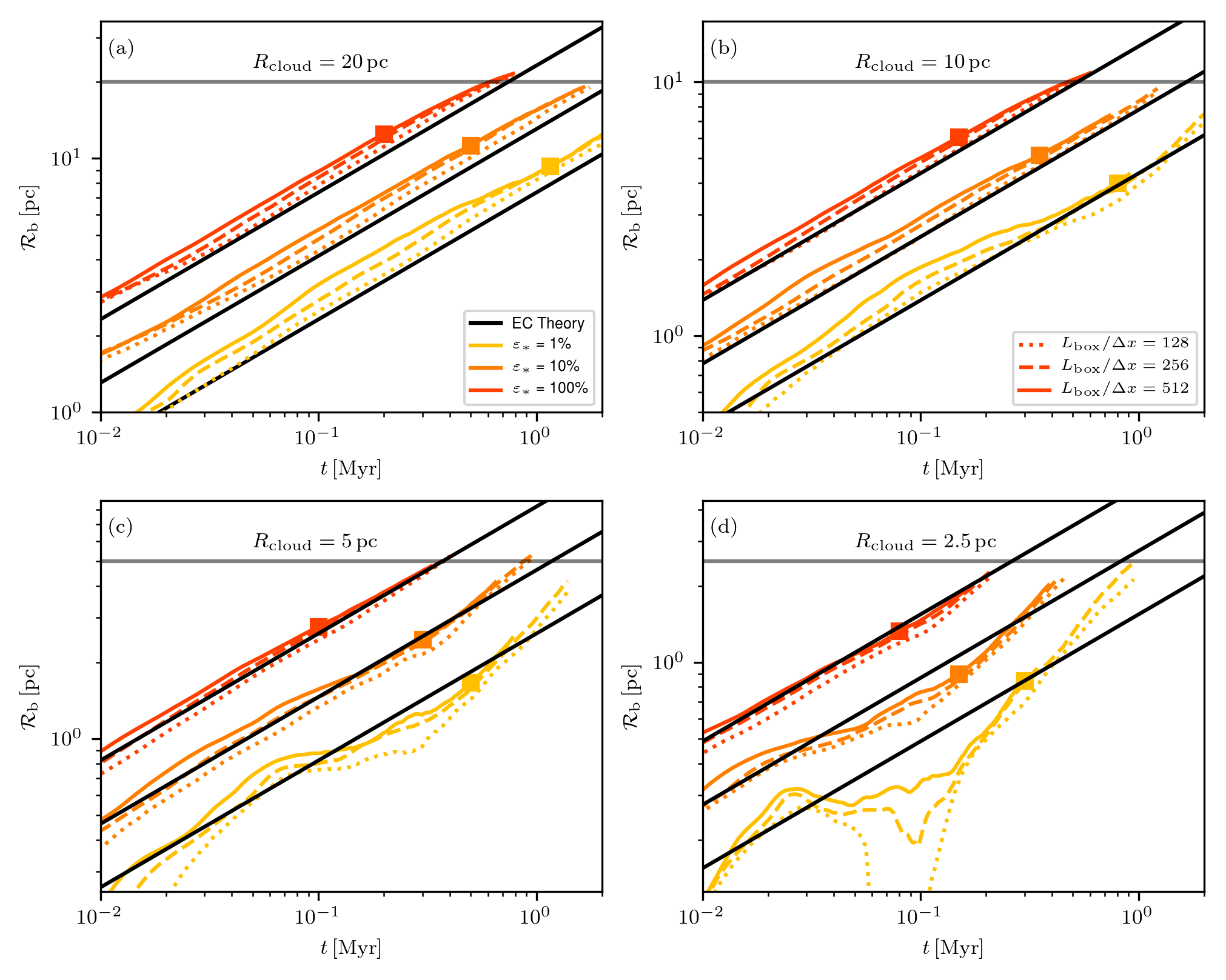}
    \caption{Evolution of the effective radius
    $\reff$ in each of our simulations for cases with cloud mass 
    $\mcloud = 10^5 M_{\odot}$.  Line styles are as in 
    \autoref{fig:pr_comp}. For each value of $\rcloud$ and 
    $\sfe$ we show (in black) the evolution prediction from the EC 
    theory (\autoref{eq:rbub_ecw}, with $\alpha_R = \alpha_p = 1$). 
    The grey horizontal lines indicate $\rcloud$, while squares indicate 
    the time when wind-contaminated gas first leaves the domain 
    ($t_{\rm esc}$). As in \autoref{fig:pr_comp}, we only show simulation 
    results up until $4t_{\rm esc}$.}
    \label{fig:radius_comp}
\end{figure*}

We note that once the bubble has reached the edge of the simulation 
domain (identified by nonzero outflow of wind-contaminated gas), we 
cannot expect the numerical solution to continue to follow the EC 
theoretical prediction.  We mark this first ``blowout'' time in 
\autoref{fig:radius_comp} with large squares (for the highest-resolution 
models).  Some time after blowout, the rate of momentum increase falls 
below $\pdot$ because a fraction of the wind exits the domain without 
interacting with cloud gas.   

\subsubsection{Effective Radius}
\label{subsubsec:meas_reff}
The effective bubble radius $\reff$ is defined from 
\autoref{eq:reff_def} based on the bubble volume, $\Vbub$. For this 
volume, we include just the hot phases of the gas, the Free Wind 
(volume $V_{\rm f}$) and the Shocked Wind (volume $V_{\rm ps}$):
\begin{equation}
    \label{eq:sim_Vbub_def}
    \Vbub \equiv V_{\rm f} + V_{\rm ps} \, .
\end{equation}

Comparisons between our simulations with $\mcloud = 10^5\, M_{\odot}$
and the prediction given by \autoref{eq:rbub_ecw} 
(with $\alpha_R=\alpha_p=1$) are given in \autoref{fig:radius_comp}. 
It is clear that the EC theory very well explains the salient features 
of the radial expansion. The only exceptions are the smallest two clouds 
for $\epsilon_*=0.01$, which is the least realistic parameter regime 
since high density regions are observed to have very high SFE 
\citep[e.g.][]{Leroy18}. These deviations are mainly caused by the 
Reynolds stress in the surrounding turbulent gas being non-negligible 
compared to the wind pressure, given the low $\sfe$.

We note in particular that the scaling $\reff \propto t^{1/2}$ 
represents the numerical results better than the steeper scaling 
$R\propto t^{3/5}$ of \citet{Weaver77} or \citet{ElBadry19} 
for pressure-driven expansion. For the majority of cases that follow 
the $\reff\propto t^{1/2}$ scaling well, the numerical result is within 
$20\%$ of \autoref{eq:rbub_ecw}. Quantitatively, the EC theory works 
especially well at higher densities (where cooling is most efficient) 
and at higher $\Lwind$ (i.e. higher $\varepsilon_*$, where $\dot p_w$ 
and $\dot{{\cal R}}_{\rm b}$ are larger, and therefore the constraint 
on $\Theta$ as given by \autoref{eq:cooling_condition} is not as stringent). 

At the time of first blowout, the mean bubble expansion rates range from 
$\sim 10-100\kms$ in the least dense cloud to $\sim 3-10 \kms$ in the 
most dense cloud, with larger velocities applying in the cases with 
higher wind power $\Lwind \propto \sfe$. Note that the first blowout 
(shown with colored squares) occurs much earlier than the time when 
$\reff(t) = \rcloud$ (shown with a horizontal grey line) due to the fractal 
nature of the bubble interface, where parts of the bubble surface are 
at much larger radii than others.

\begin{figure*}
    \centering
    \includegraphics{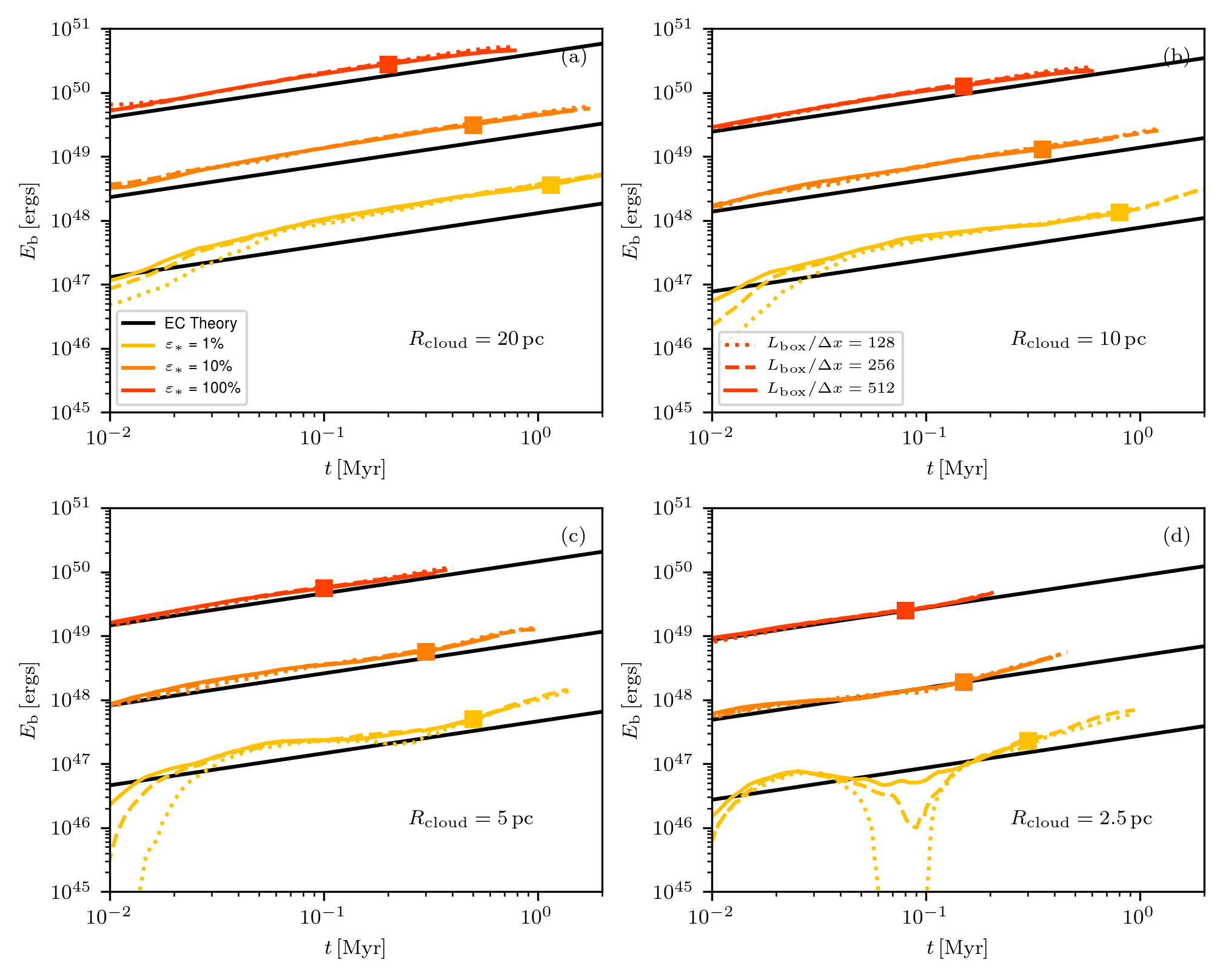}
    \caption{Evolution of the total energy in the interior of the 
    bubble in each of our simulations for cases with cloud mass 
    $\mcloud = 10^5 M_{\odot}$. Line styles are as in 
    \autoref{fig:pr_comp}. For each value of $\rcloud$ and 
    $\sfe$ we show in black the energy given by
    \autoref{eq:ebub_relation} (taking $\ratr=\alpha_p=\alpha_R=1$ and using \autoref{eq:rbub_ecw}).}
    \label{fig:energy_comp}
\end{figure*}

\subsubsection{Bubble Interior Energy}
\label{subsubsec:meas_Eb}
In keeping with the definition of the bubble interior given 
by \autoref{eq:sim_Vbub_def}, we measure the bubble's interior's 
energy using
\begin{equation}
    \label{eq:sim_ebub}
    \Ebub \equiv E_{{\rm f},{\rm kin}} + E_{{\rm f},{\rm thm}}
    + E_{{\rm ps},{\rm kin}} + E_{{\rm ps},{\rm thm}}
\end{equation}
where the ``${\rm kin}$'' and ``${\rm thm}$'' subscripts refer to 
kinetic and thermal energy respectively. We note that the 
thermal energy within the free wind ($E_{{\rm f},{\rm thm}}$) 
is negligible compared to its kinetic energy, but we include it 
here for completeness.

\autoref{fig:energy_comp} compares the energy  measurement from 
the simulations with the prediction given by \autoref{eq:ebub_relation}, employing \autoref{eq:rbub_ecw} and taking $\ratr=\alpha_p = \alpha_R = 1$. When comparing the theory and simulations,
we see similar trends as those observed in the evolution of bubble radii 
and momenta: the EC theory is most accurate at higher $\sfe$ and higher 
density $\rhobar$.

We also note that given the agreement with theoretical predictions for 
the shell momentum (as evidenced by \autoref{fig:pr_comp}) and the bubble 
internal energy (as evidenced by \autoref{fig:energy_comp}), we expect the
shell's radial kinetic energy to be half that of the bubble's interior energy, 
as predicted in Paper I (see Equation 23 there). This is indeed the case.

\begin{figure*}
    \centering
    \includegraphics{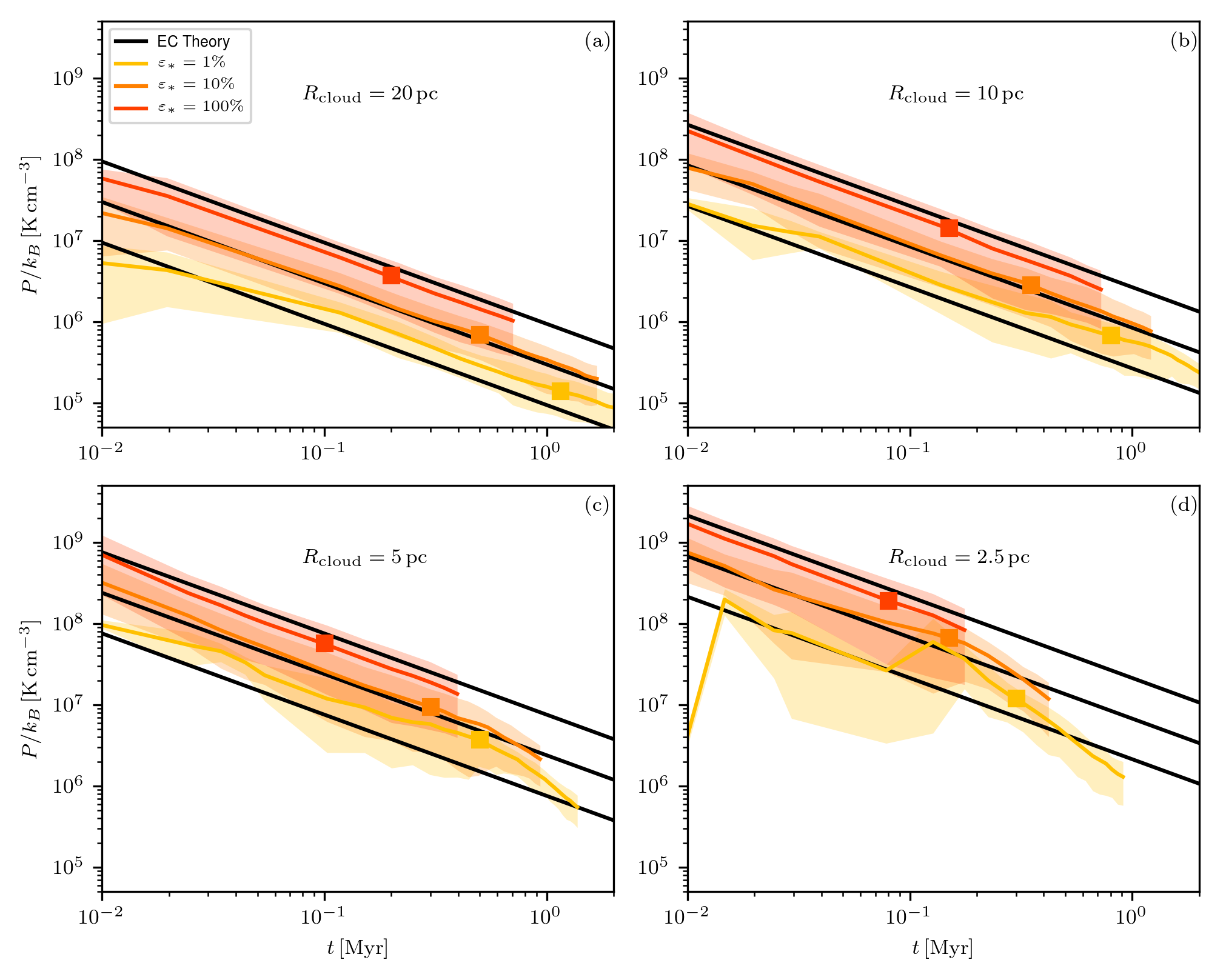}
    \caption{Evolution of the pressure in the shocked wind gas in 
    cases with cloud mass $\mcloud = 10^5 M_{\odot}$. All panels show the 
    evolution of the volume averaged mean pressure (solid colored lines) 
    along with shading the $16^{\rm th}$ to $84^{\rm th}$ percentiles of 
    the distribution of pressures in post-shock gas (colored shaded regions). 
    We show the theoretical prediction (\autoref{eq:pressure}, with $\alpha_p=1$) in 
    black. The time at which wind first escapes the domain is indicated 
    by a colored square.}
    \label{fig:pressure}
\end{figure*}

\subsubsection{Hot Gas Pressure}
\label{subsubsec:pressure}

Finally, we compare the prediction for the pressure in the shocked wind 
gas given by \autoref{eq:pressure} and that measured in our simulations. 
We measure the pressure by selecting the shocked wind gas 
(conditions specified in \autoref{tab:phase_defs}) in simulation snapshots 
and calculating the volume-averaged mean thermal pressure.  We use the 
$\Lbox/\delx = 256$ simulations for these comparisons as the pressure is 
computed directly from snapshots, which are taken at higher cadence in the 
lower resolution simulations. We also compute the $16^{\rm th}$, $50^{\rm th}$, 
and $84^{\rm th}$ percentiles of the distribution of pressures in this gas.

In \autoref{fig:pressure} we compare the evolution of the pressure with the 
prediction of \autoref{eq:pressure}. For this comparison, we use 
$\alpha_p=\alpha_R=1$, and we see that the theoretical prediction closely 
matches the evolution in the simulations.  As we discuss below, the simulations 
actually have $\alpha_p > 1$. From \autoref{eq:pressure} this would tend to 
increase the theoretical prediction (black curves in \autoref{fig:pressure}), 
bringing the prediction closer to the `observed' simulation value for the low 
$\sfe$ cases (which have the largest $\alpha_p$), but further away from 
agreement for high $\sfe$ cases. The missing component here is the obliquity of 
the shock surface: the more oblique the shock the less thermalised the shocked 
wind becomes, hence lowering the pressure (discussed at the end of Appendix 
A in Paper I). As is clear from \autoref{fig:example_slice_R20} and 
\autoref{fig:example_slice_R2p5}, the high $\sfe$ winds have more oblique 
shocks, explaining this discrepancy.

\begin{figure*}
    \centering
    \includegraphics{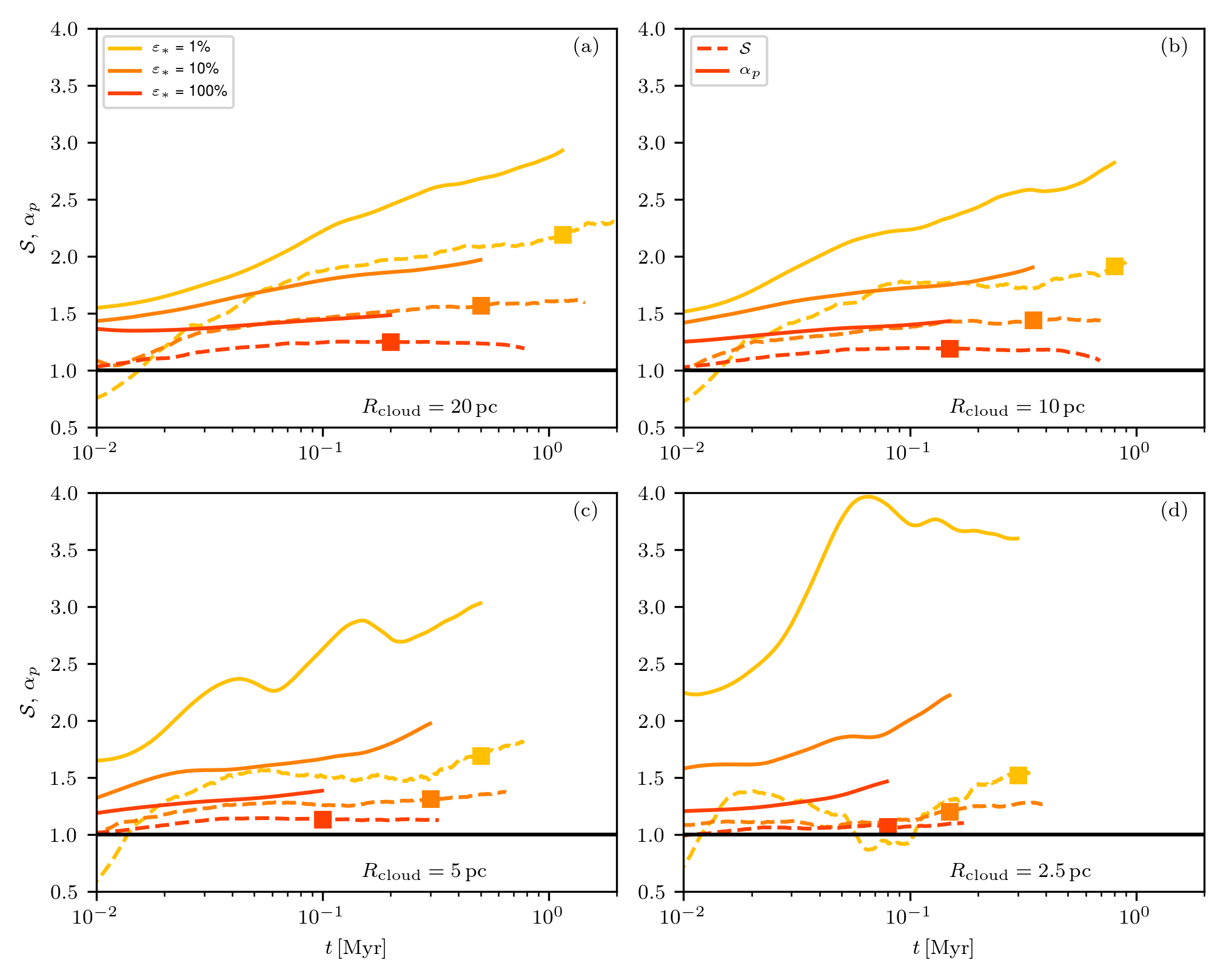}
    \caption{Evolution of the momentum ($\alpha_p$) and energy 
    ($\ratr$) enhancement factors shown as solid and dashed lines 
    respectively. The evolution of $\alpha_p$ is only shown up to the point 
    of breakout, which is indicated with colored squares on the curves for 
    $\ratr$. We indicate the `no enhancement' case with the horizontal 
    black line. We show the highest resolution version of all simulations 
    with $\mcloud=10^5 \, M_{\odot}$.}
    \label{fig:energy_ratio}
\end{figure*}

\begin{figure*}
    \centering
    \includegraphics{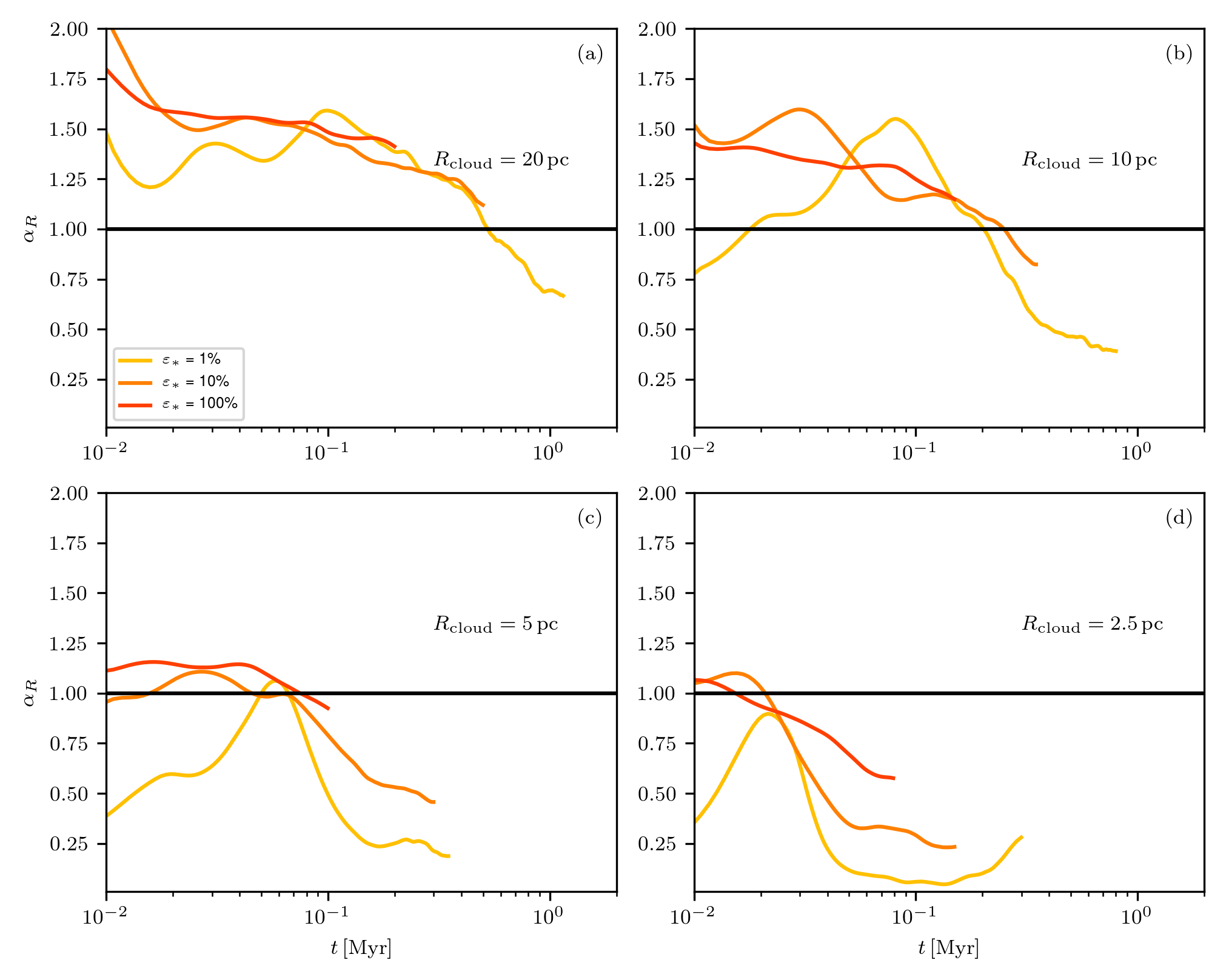}
    \caption{Evolution of the dimensionless parameter $\alpha_R$. 
    Line colors are associated with simulations of different $\sfe$.  
    We use the highest resolution version of all simulations with 
    $\mcloud=10^5\, M_{\odot}$ and only show the evolution up to the 
    point of breakout (square markers in \autoref{fig:pr_comp} 
    and others).}
    \label{fig:alphaR}
\end{figure*}

\begin{figure*}
    \centering
    \includegraphics{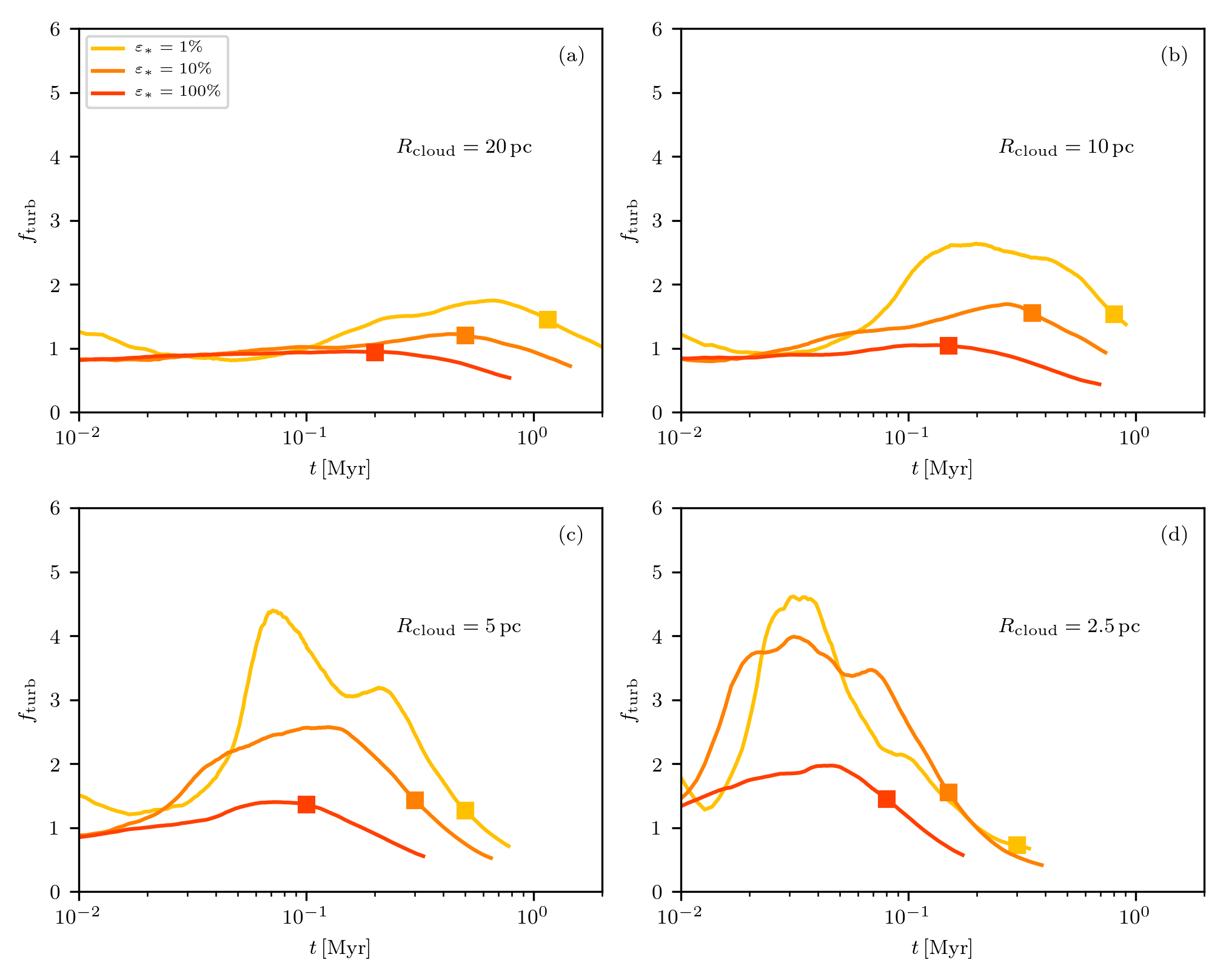}
    \caption{Evolution of the ratio of turbulent kinetic energy 
    to kinetic energy in radial motion, for shell gas. We show the 
    highest resolution version of all simulations with 
    $\mcloud=10^5 \, M_{\odot}$. The square symbols mark the point 
    at which wind-polluted mass begins to leave the domain.}
    \label{fig:fturb}
\end{figure*}

\subsection{Dimensionless Parameters}
\label{subsec:params}

In the above comparisons we explicitly set the dimensionless, 
order-unity parameters of our theory equal to unity. However, the 
values and time evolution of these parameters provides interesting 
insight into the validity of the assumptions in the EC theory for 
different regimes (see \autoref{subsec:bubble_evol}). 
As is explained in Appendix A of Paper I and summarized in 
\autoref{sec:theory}, some of these parameters are interdependent.

The first dimensionless parameter is $\alpha_p$, which is defined 
as the rate of momentum input to the surrounding medium divided by 
the rate of momentum injection by the wind. Specifically, we measure
\begin{equation}
    \label{eq:meas_alphap}
    \alpha_p = \frac{p_{\rm r, b}}{\pdot t}
\end{equation}
where $\pr$ is as measured in \autoref{eq:sim_pr_def}. 
This measurement of $\alpha_p$ is shown as solid lines in 
\autoref{fig:energy_ratio}.

Another quantity in the EC theory is the energy enhancement factor 
$\ratr$, defined in \autoref{eq:ebub_relation}. As explored in depth 
in Appendix A of Paper I, $\ratr$ and $\alpha_p$ are expected to track 
each other (with $\ratr \approx \alpha_p$) because both are associated 
with buildup of energy in the bubble interior. For the same reason, we 
expect $\ratr$ to be higher when the shocked wind makes up a larger 
portion of the bubble volume, i.e. larger $\reff/\rfree$.  We also 
expect larger $\ratr$ when shock surfaces are less oblique. In order to
directly compare $\ratr$ with $\alpha_p$, we show in 
\autoref{fig:energy_ratio} a measured value of $\ratr$ as dashed lines. 

The measured value of $\ratr$ is taken as the ratio of the bubble 
energy in the simulations, measured according to \autoref{eq:sim_ebub}, to 
$\pdot \reff/2$, where $\pdot$ is taken as the fixed wind momentum 
input rate and $\reff$ is computed from the simulations as detailed 
in \autoref{subsubsec:meas_reff}.

The next dimensionless parameter we introduce is $\alpha_R$, which 
encodes geometric factors. Using \autoref{eq:rbub_ecw}, 
\begin{equation}
    \label{eq:alphaR_def}
    \alpha_R = \frac{\reff^4}{\alpha_p} 
    \left(\frac{3}{2\pi} \frac{\pdot t^2}{\rhobar} \right)^{-1} \, .
\end{equation}
We evaluate this using the measured $\reff$ (see \autoref{subsubsec:meas_reff}) 
and $\alpha_p$ (\autoref{eq:meas_alphap}). We note that we have no explicit 
theoretical prediction for $\alpha_R$ other than expecting it to be near unity.
\autoref{fig:alphaR} shows the measured values of $\alpha_R$ over time. 
We see that $\alpha_R$ remains quite close to unity, remaining in the 
range $0.5-1.5$ for the majority of the evolution. There is also some 
indication that $\alpha_R$ takes on lower values in the higher density 
clouds.

The \textit{radial} kinetic energy of the shell in the EC theory is 
determined by momentum input from the wind, but there is no prediction 
for the kinetic energy in non-radial, turbulent motion. Given that 
the theory relies on efficient cooling facilitated through a turbulent 
interface, we expect a significant fraction of the kinetic energy in the 
shell (and around the shell-bubble interface) to be in turbulent motion. 
The simplest way to quantify this is in terms of the $f_{\rm turb}$ 
parameter defined in \autoref{eq:fturb_def}.

To measure this parameter we write the total radial kinetic energy 
in the shell as 
\begin{equation}
    \label{eq:sim_ekr_def}
    E_{r,{\rm sh}} \equiv \sum_{j\in {\rm i,w,u,c}} 
    \frac{p_{r, j}^2}{2 M_{j}}
\end{equation}
where the sum is performed over the indicated phases, and as in 
\autoref{eq:sim_pr_def} we sum over zones where $\fwind > 10^{-4}$.
The total kinetic energy is measured in an analogous way.

The ratio $f_{\rm turb}=E_{\rm tot,sh}/E_{\rm r,sh} -1 $ is displayed 
in \autoref{fig:fturb}. Surprisingly, for much of the bubble evolution 
and most of the parameter space, the energy in turbulent motion is at 
least as large as the energy in radial motion, with $f_{\rm turb} \approx 1$
(equipartition between radial and turbulent motion) over much of the 
parameter space. This fact emphasizes how important the turbulence is 
for the evolution of the bubble, especially the turbulence driven by the 
wind. We observe that the more powerful winds instill a larger fraction 
of their kinetic energy in radial motion.

In all simulations the fraction of turbulent kinetic energy in the shell 
increases in time until seemingly reaching a set value before decreasing 
again. As we will see upon further inspection of the turbulent motion 
below, this reflects a saturation of turbulence as more and more radial 
energy is provided to the wind.

\begin{figure*}
    \centering
    \includegraphics{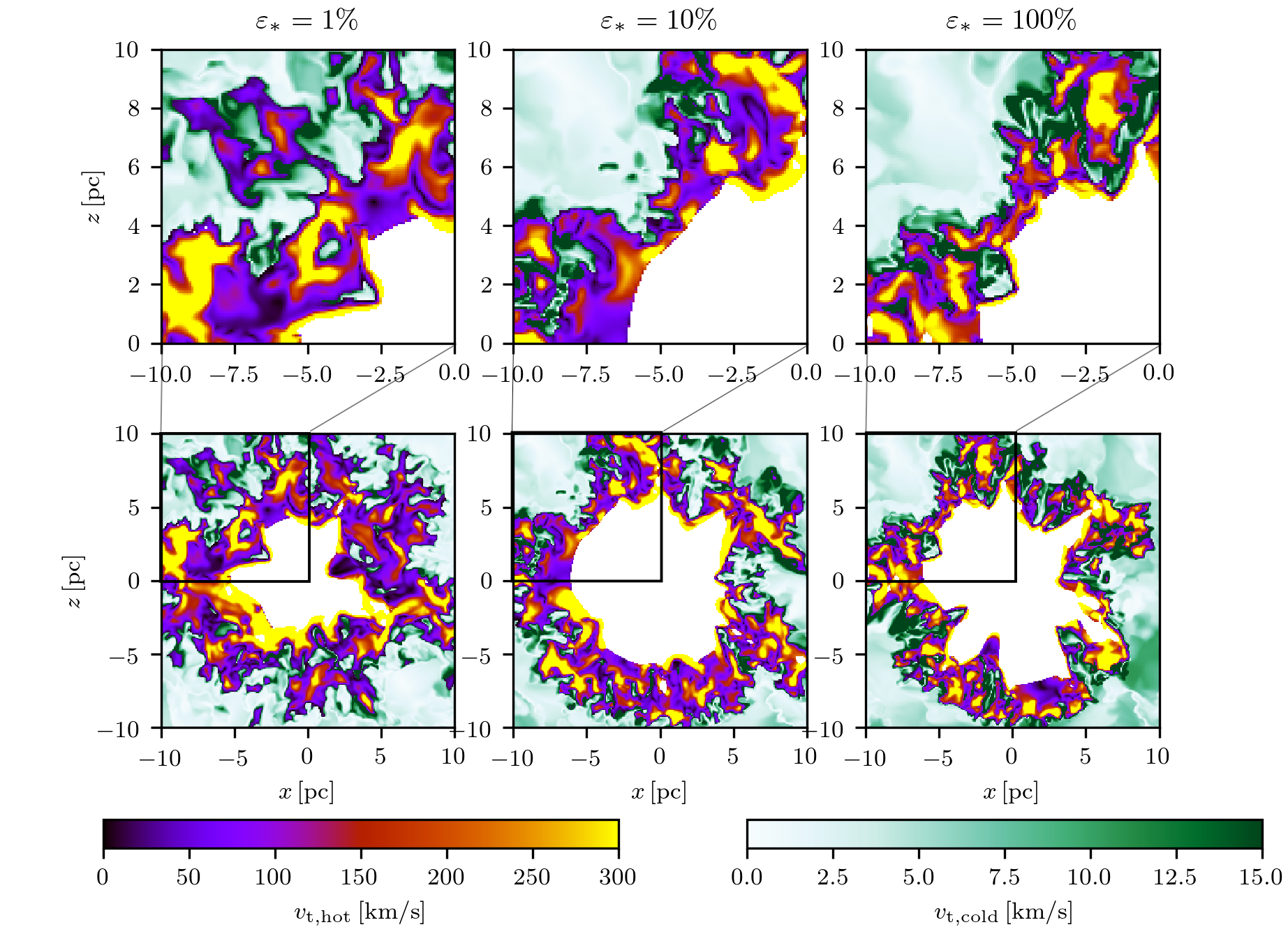}
    \caption{Magnitude of the tangential velocity in snapshots of the 
    $\rcloud = 20 \, {\rm pc}$ models at the time in each simulation when 
    $\reff = \rcloud/3$.  The separate columns (left to right) are the 
    $\sfe = 1\%,\, 10\%,\, $and $100\%$ cases. The hot gas (Shocked Stellar 
    Wind and Ionized Gas) is shown with the purple-to-yellow color scale 
    while the cold gas (Warm, Thermally Unstable, and Cold Gas) is shown 
    with the green color scale. The top row of panels is a zoom-in of the 
    bottom row, which itself zooms in on just the bubble region, to 
    highlight the structure of turbulence in the bubble/shell interface.}
    \label{fig:vt_show}
\end{figure*}

\begin{figure*}
    \centering
    \includegraphics{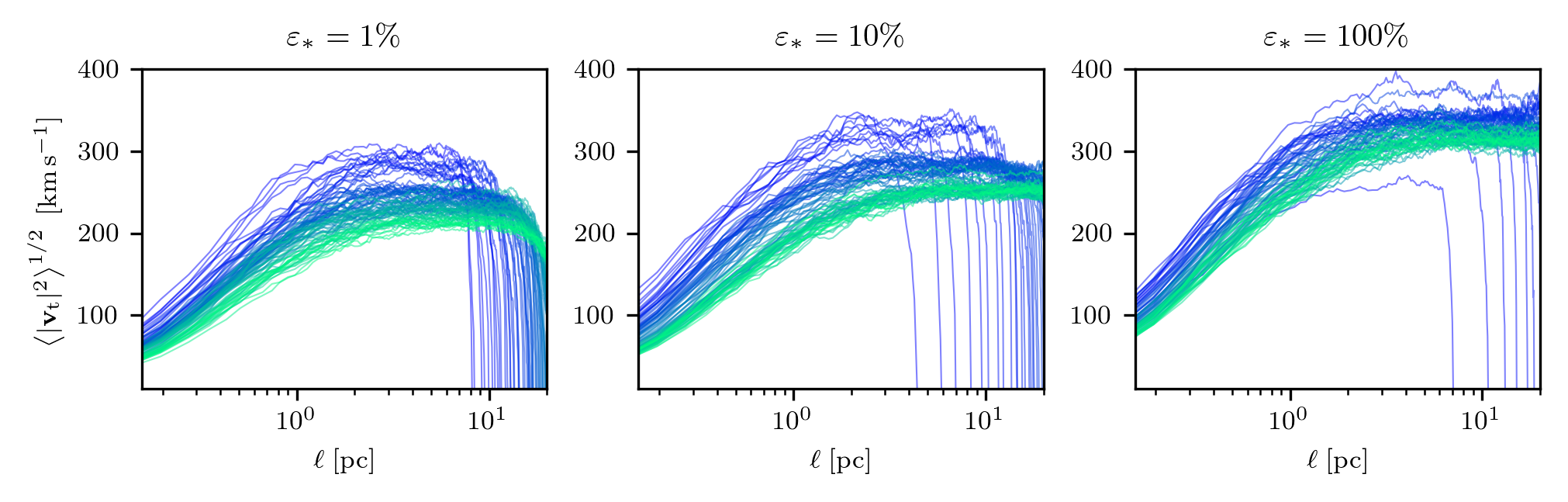}
    \caption{Temporal evolution of the structure function of 
    turbulent velocities in the hot gas (Shocked Stellar Wind and Ionized Gas)
    in the first $1\, {\rm Myr}$ of the simulation for the 
    $\rcloud=20\, {\rm pc}$, $\mcloud= 10^5 M_{\odot}$ simulations. The 
    separate panels (left to right) are the $\sfe = 1\%,\, 10\%,\, $and $100\%$ 
    cases. Temporal evolution is indicated by the color of the lines in each 
    panel, moving from early times (dark blue) to the late times (bright green) 
    in increments of $10^4\, {\rm yrs}$. The characteristic velocity scale of 
    the turbulence decreases in time and varies by at most a factor of 2 
    over the course of the simulation. The characteristic spatial scale of 
    the structure function increases over time. Turbulent velocities are 
    overall higher for higher wind power ($\Lwind \propto \sfe$) models.}
    \label{fig:vt_struct_ev}
\end{figure*}

\begin{figure*}
    \centering
    \includegraphics{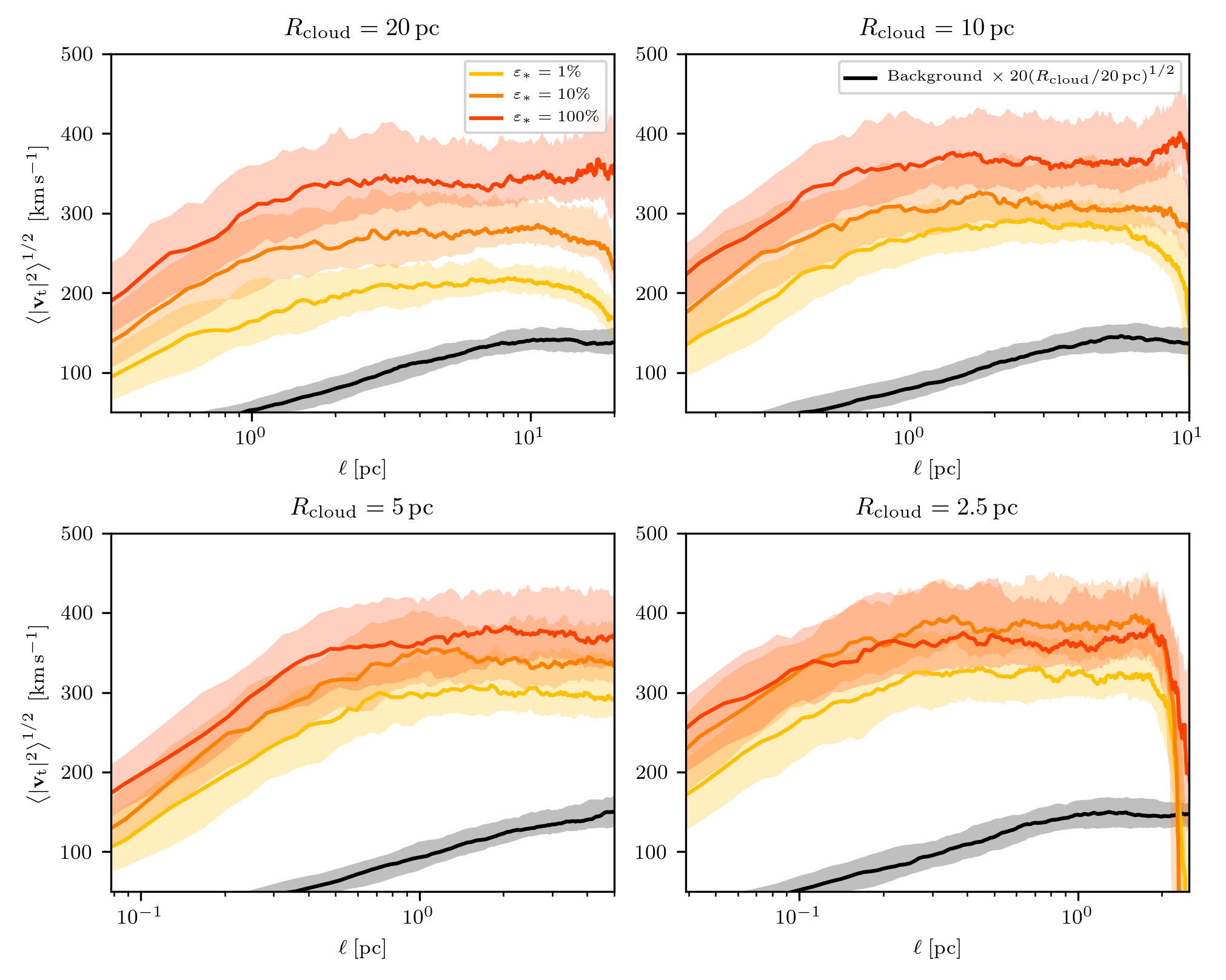}
    \caption{The structure function of turbulent velocities in the hot 
    gas (Shocked Stellar Wind and Ionized Gas) when $\reff = \rcloud/3$. 
    These are measured in the highest resolution versions of all 
    simulations with $\mcloud= 10^5 M_{\odot}$ and we show the simulations 
    with $\sfe = 1\%, 10\%, 100\%$ in yellow, orange, and red respectively. 
    For each panel we show scales from $4\delx$ to $\rcloud$ (solid lines) 
    as well as the 25$^{\rm th}$ to 75$^{\rm th}$ percentiles of structure 
    function measurements over the ensembles of 200 samples from the hot 
    gas.  In each panel we show in black the structure function of the 
    background turbulent gas scaled up by a factor $20(\rcloud/20\pc)^{1/2}$. 
    The velocity scale of the turbulence is set by the strength of the wind 
    (represented here by $\sfe$) while the characteristic length scale of 
    the turbulence is likely set by the bubble size and turbulent 
    density structure of the background.
    }
    \label{fig:structure_functions}
\end{figure*}

\subsection{Turbulent Structure Function}
\label{subsec:turbulent_struct}
Both the amplitude and spectral shape of turbulence are important to 
the mixing and cooling that dictate the bubble evolution.  These are 
quantified via the turbulent structure function.

Following the theory outlined in Paper I, we 
particularly wish to measure the turbulence in the hot gas near the 
interface. To this end, we select gas with $T>2\times 10^4\, {\rm K}$ 
and $v_r < \vw /2$ (the Ionized Gas and the Shocked Stellar Wind). In 
order to isolate the turbulent motion from the bulk radial outflow in 
this gas, we apply our structure function analysis only to the non-radial 
components of the velocity field.  We will refer to this proxy for the 
turbulent velocity field (which is really the tangential velocity field) 
as $\mathbf{u}_t$ defined as
\begin{equation}
    \mathbf{u}_t \equiv \mathbf{v} 
    - \left(\mathbf{v}\cdot \mathbf{\hat{r}}\right) \mathbf{\hat{r}}  \, .
\end{equation}
where $\mathbf{v}$ is the full velocity field. Of course, there are 
also turbulent contributions to the radial motion, but directly quantifying 
this is problematic due to contamination by the strong background radial 
flow. \autoref{fig:vt_show} shows example snapshots of the magnitude of 
$\mathbf{u}_t$ from the $\rcloud=20\pc$ model (in the purple to yellow
color scheme). 

To measure the turbulent structure function, we randomly select a cell $i$ 
from the region under consideration. We then compute two histograms. The first 
histogram is the number of cells at position $j$ with a spatial separation 
$\ell = |\mathbf{x}_j - \mathbf{x}_i|$ from cell $i$, binned by width $\delx$. 
The second histogram is as above, except now the histogram is weighted 
by the square of the difference in the tangential velocity field between 
positions $i$ and $j$:
\begin{equation}
    |\delta \mathbf{u}_t|^2 \equiv 
    \left| \mathbf{u}_t(\mathbf{x}_i) 
    - \mathbf{u}_t (\mathbf{x}_j) \right|^2 \, .
\end{equation}
Dividing the second histogram by the first and taking a square root 
gives us a single sampling of the root-mean-square velocity offset 
as a function of separation scale $\ell$. In order to account for the 
(unmeasured) radial component of the true turbulent velocity field, we 
additionally multiply by $\sqrt{3/2}$ (assuming that the turbulence is 
isotropic). We repeat the above process for 200 cells randomly selected 
from the Ionized Gas and Shocked Stellar wind. 

With this factor of $\sqrt{3/2}$ we refer to the measured structure 
function as $\strfunc$. With these 200 samples of $\strfunc$ we take 
the median value in each radial bin (over the ensemble of samples) to represent 
the structure function. We also calculate the 25$^{\rm th}$ to 75$^{\rm th}$ 
percentiles over these 200 samples to quantify the width of the distribution of 
velocity deviations at each scale.

As the calculation described above is quite computationally 
expensive, we only calculate the full time evolution of the 
structure function for a subset of our simulations 
($\rcloud = 20 \, {\rm pc}$, $\mcloud = 10^5 \, M_{\odot}$, 
and $\Lbox/\delx = 512$). The results of these calculations 
are illustrated in \autoref{fig:vt_struct_ev}, where the 
turbulent structure function is shown in time increments of 
$10^4\, {\rm yrs}$ varying from early times (shown in dark blue)
to late times  (shown in bright green).  The beginning time is 
$0.01\, {\rm Myr}$ and the ending times are $2.23,$ $1.21,$ and 
$0.88\, {\rm Myr}$ for the $\sfe = 1\%,$ 10\%, and 100\% cases 
respectively. At all star formation efficiencies the characteristic 
velocity scale of the turbulence decreases in time but only by at most a 
factor of two.  In comparison, the shell velocity decreases by a factor 
of 10 over the same time interval.

For the rest of our  simulations with $\mcloud = 10^5 \, M_{\odot}$ 
and $\Lbox/\delx=512$, we measure the structure function of the turbulence at 
the individual time when $\reff = \rcloud/3$. We show these structure 
functions, along with the 25$^{\rm th}$ to 75$^{\rm th}$ percentiles 
of velocity deviations, in \autoref{fig:structure_functions}. To 
provide a reference for the background turbulence, we calculate in 
the same way the structure function of the turbulent background in 
the full simulation volume just before the initiating of the wind. 
This background turbulence has a much lower velocity than the turbulence 
in the hot gas, so 
we scale the background measurements by a factor $20(\rcloud/20\pc)^{1/2}$;
this scale choice is purely for plotting purposes.
This scaling also adjusts for the differences in initial turbulent 
velocity scales $v_t \propto \rcloud^{-1/2}$ amongst the different size 
clouds (as laid out in \autoref{tab:sim_params}) so that all background 
structure functions appear essentially identical.

It is clear from \autoref{fig:structure_functions} that the velocity 
scale of the turbulence is determined primarily by the strength 
of the wind, which is represented here by $\sfe$. However, the 
characteristic spatial scale of the turbulence, which is where the 
structure function flattens out, is consistently $\sim \rcloud/10$, 
suggesting that this scale may be set by the size of the bubble and/or 
the spatial scale of the background turbulence. This variation in 
turbulent velocity magnitude with $\sfe$ could be explained by the 
more oblique shocks in the higher $\sfe$ cases: when a shock is 
more oblique the post-shock flow will have a larger speed.

Along those lines, we note that for the 
$\sfe = 100\%$ case in the  $\rcloud=2.5\; {\rm pc}$ cloud, the 
amplitude of the turbulent velocity is lower than expected. This is 
because the $v_r < \vw/2$ restriction cuts out much of the shocked 
wind due to the strongly-oblique shocks in this case 
(see \autoref{fig:example_slice_R2p5}).

\begin{figure*}
    \centering
    \includegraphics{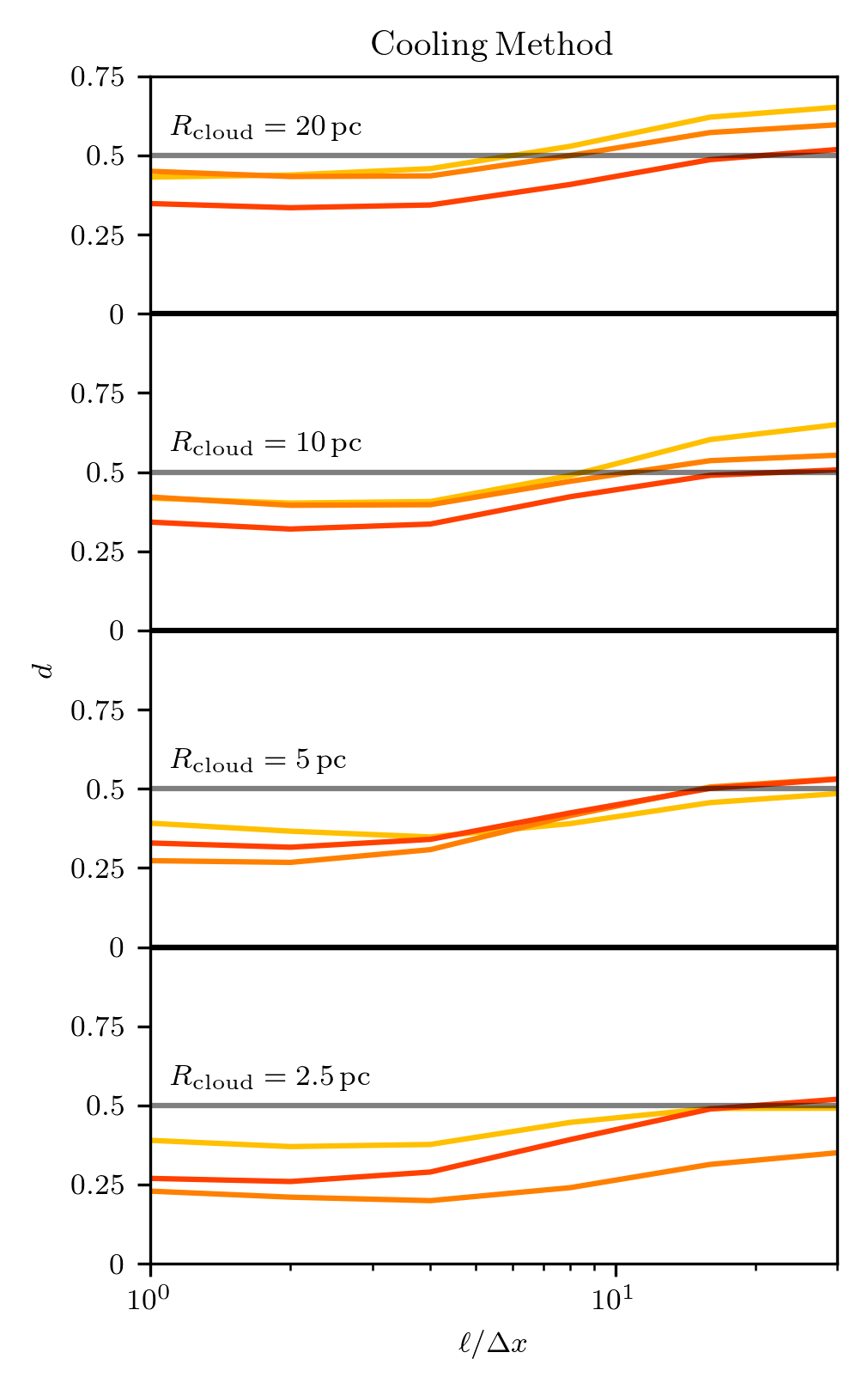}
    \includegraphics{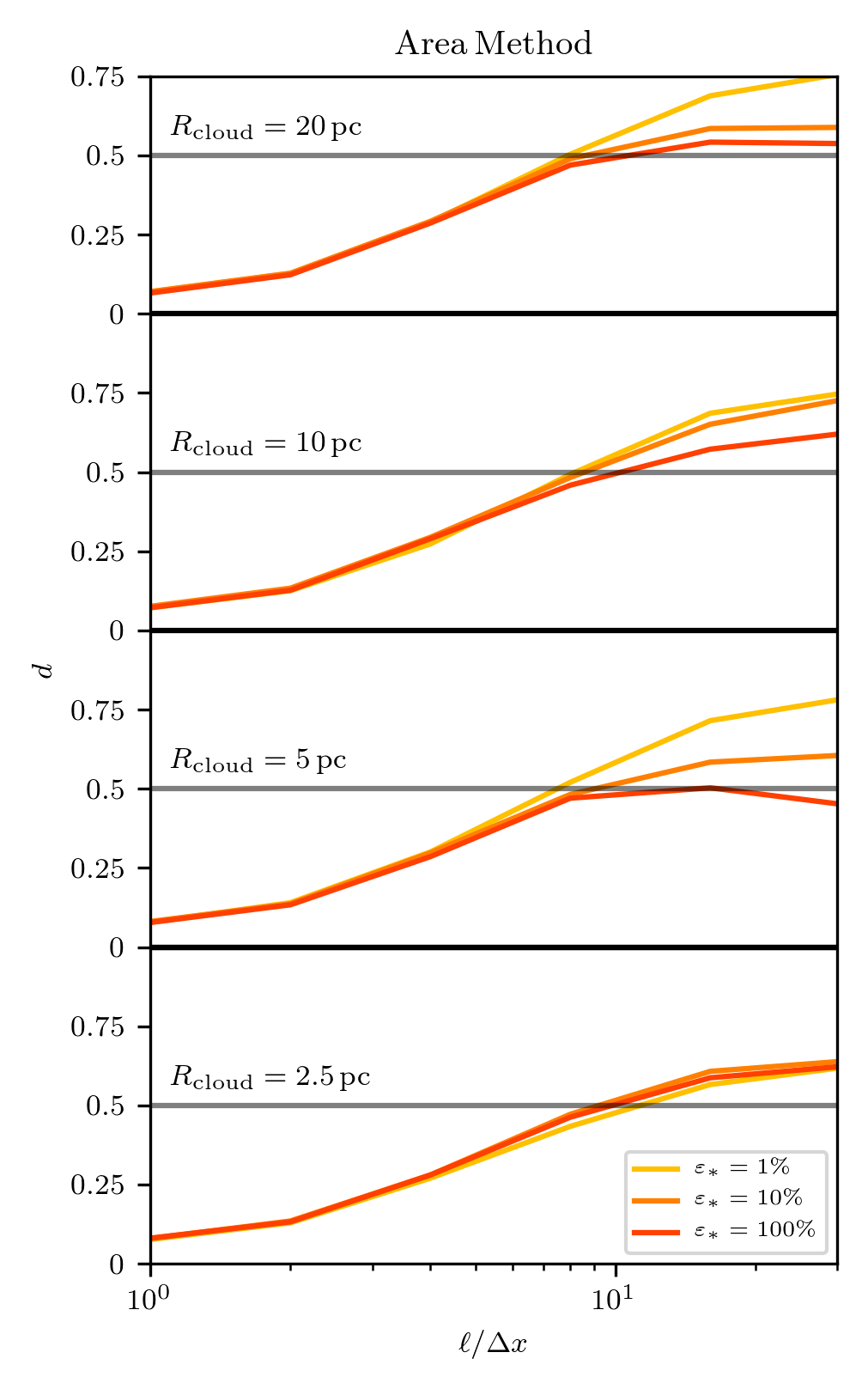}
    \caption{
    The excess fractal dimension of the bubble surface 
    as a function of scale using the two measurement techniques 
    described in the text (see  \autoref{subsubsec:cooling_fractal} 
    and \autoref{subsubsec:area_fractal}). \textit{Left panel:} The 
    excess fractal dimension of the bubble surface measured using a 
    box-counting technique; boxes at different scales mark where cooling is taking place. \textit{Right Panel}: The excess fractal dimension measured 
    based on the logarithmic scaling of the area of an iso-temperature 
    surface ($T=10^5\, {\rm K}$ for all curves shown here) when measured 
    on different scales.
    }
    \label{fig:fractal}
\end{figure*}

\subsection{Fractal Structure of Interface}\
\label{subsec:frac_cooling_analysis}

In this section we explore the fractal nature of the shell-bubble interface. 
These results also inform the discussion of turbulence-induced mixing and 
cooling in Paper I.

\subsubsection{Fractal Dimension from Cooling}
\label{subsubsec:cooling_fractal}
We measure the fractal dimension of the bubble surface in two different 
ways. Since the details of the fractal structure of the bubble/shell 
interface affect the cooling, it makes sense to measure the fractal 
dimension using the cooling itself. To that end we determine the fractal 
dimension using a box-counting or Minkowski-Bouligand algorithm 
\citep{Schroeder91} where membership in or out of the fractal is determined 
based on cooling.\footnote{The measurement technique we adopt was suggested 
to the authors by Drummond Fielding.} To maximize dynamic range we use 
simulations with $\Lbox/\delx = 512$.  To be consistent with the turbulent 
structure function measurements of \autoref{subsec:turbulent_struct}, we 
analyze the fractal structure at the time when $\reff = \rcloud/3$.

We first choose a fraction $\fcool \in (0,1)$. For a given time snapshot 
we measure the total amount of cooling, $\ecool$. We denote 
$\dot{e}_{{\rm cool}, i}$ as the cooling rate in the $i^{\rm th}$ 
hydrodynamical cell, for a  sorted list starting with the maximal cooling 
rate. We then compute the cumulative sum up to $I$ such that 
$\sum_{i=0}^{I} \dot{e}_{{\rm cool}, i} \approx \fcool \ecool$. The count $I$ of 
cells contributing to the specified cooling fraction at the minimum scale 
$\ell = \delx$ (the resolution of the simulation) is hereafter denoted by 
$I(\Delta x)$. 

We next reduce the size of the simulation grid by a factor of two 
in each direction, so that merged cells have length $\ell=2\Delta x$, and 
the cooling of adjacent cells is combined. We repeat the above 
sort-and-accumulate process to measure the number of cells $I(2\Delta x)$ 
needed for $\sum_{i=0}^{I(2\Delta x)}\dot{e}_{{\rm cool}, i}\approx\fcool\ecool$.
We iteratively repeat this process for $\ell/\delx =  1,\, 2,\, 4,\, 8,\, 16,\,$ 
and $32$ and use these measurements to calculate the fractal dimension as 
\begin{equation}
    D = d + 2 = -\frac{\diff \log I(\ell)}{\diff \log \ell} \, .
\end{equation}
This quantity will vary with scale, $\ell$.  However, we expect $d$ to take on 
a roughly constant value on large scales, dropping  below this at sufficiently 
small scale due either to limited numerical resolution or to some physical 
dissipation scale such as the cooling length, $\lcool$. In practice the former 
is more relevant as the cooling length is usually below our resolution limit.

At sufficiently large $\fcool$, $d$ 
is close to zero on large scales and becomes very large (based on our fitting 
method) on small scales. This is because $I(\fcool,\ell)$ becomes a very steep 
function of $\fcool$ when transitioning beyond the concentrated cooling in the 
interface boundary layer.  Above a certain $\fcool$, essentially the whole 
domain is included, implying that the true $d\rightarrow 1$ at small $\ell$. 

We believe that the most physical $d$ to associate with the fractal structure 
of the cooling interface is the value at large scales for the largest value 
of $\fcool$ that does not exhibit the discontinuous behaviour explained above.
On the left hand side of \autoref{fig:fractal}, we therefore show the dependence 
of $d$ on scale for the largest choice of $\fcool$ that does not result in this 
discontinuous behavior.  This choice of $\fcool$ varies between simulations.

The scale dependent nature of the ``excess'' fractal dimension $d$ is evident 
in \autoref{fig:fractal}, where $d$ decreases at small scales. We see that $d$ 
tends to be larger at smaller values of $\sfe$ (this is  clearest in the 
larger/lower density clouds). In part, this may be because turbulence 
at the bubble-shell interface, which is responsible for creating the fractal 
structure, makes up a much larger fraction of the kinetic energy at low $\sfe$, 
as was shown in \autoref{fig:fturb}.

Overall, this measurement method  suggests a fractal dimension $d\approx 1/2$ 
in most simulations, with larger values (up to $\sim 0.6-0.7$) in models with 
very low $\sfe$ and large $\rcloud$. 

\subsubsection{Fractal Dimension from Iso-Temperature Surfaces}
\label{subsubsec:area_fractal}
Our second method of measuring the fractal dimension, paralleling  
\citet{FieldingFractal20}, uses the area of iso-temperature surfaces 
for temperatures that should be characteristic of the interface between 
the wind bubble and the shell. Specifically, we measure the area of 
iso-temperature surfaces  for $T = 10^4$, $10^{4.5}$, $10^5$, 
$10^{5.5},$ and $10^6 \, {\rm K}$.  We use the \texttt{marching\_cubes} 
algorithm from the \texttt{scikit-image} package to measure these surface 
areas, using step sizes $\ell/\delx = 1,\, 2,\, 4,\, 8,\, 16,\,$and $32$. 

The excess dimension $d$ is then given by the logarithmic derivative of 
the area with respect to the scale on which the area is measured
\begin{equation}
    \label{eq:frac_d_marchingcubes}
    d = -\frac{\diff \log \Abub(\ell)}{\diff \log \ell} \, .
\end{equation}
As with the previous method, this value should be a constant over a large 
range of scales, decreasing at some small scale due to numerical or 
physical dissipation (usually the former). Our measurements are made at the 
time when $\reff = \rcloud/3$.  The scale-dependent results are shown 
at the right side of \autoref{fig:fractal}.
For the sake of brevity we only show the results for the iso-temperature 
surfaces measured at $T=10^5\, {\rm K}$ for each simulation.

\autoref{fig:fractal} clearly shows that this measurement technique 
consistently finds $d \rightarrow 0$ at small scale, as expected. 
We also found (not shown)
that the value of $d$ on large scales 
depends on temperature: the higher temperature surfaces have more fractal 
(higher $d$) structures than the lower temperature surfaces, with the effect 
most extreme in the smallest-$\sfe$, largest-$\rcloud$ models. On large scales, 
the range of $d$ from this technique falls roughly within $d\sim 1/2-3/4$.

It is also interesting to note that at all cloud sizes, the bubble 
surface tends to become less fractal at higher wind luminosity (or 
$\sfe$), as seen in the previous measurement technique, but 
now to a higher degree.
Physically, the more-fractal structure in the lower-$\sfe$ models at 
large scales may potentially be explained by their greater buildup of hot 
gas (higher $\ratr$), which can more effectively create ``fingers'' of 
high-temperature gas within the lower-density parts of the turbulent cloud 
at large scales.

\begin{figure*}
    \centering
    \includegraphics{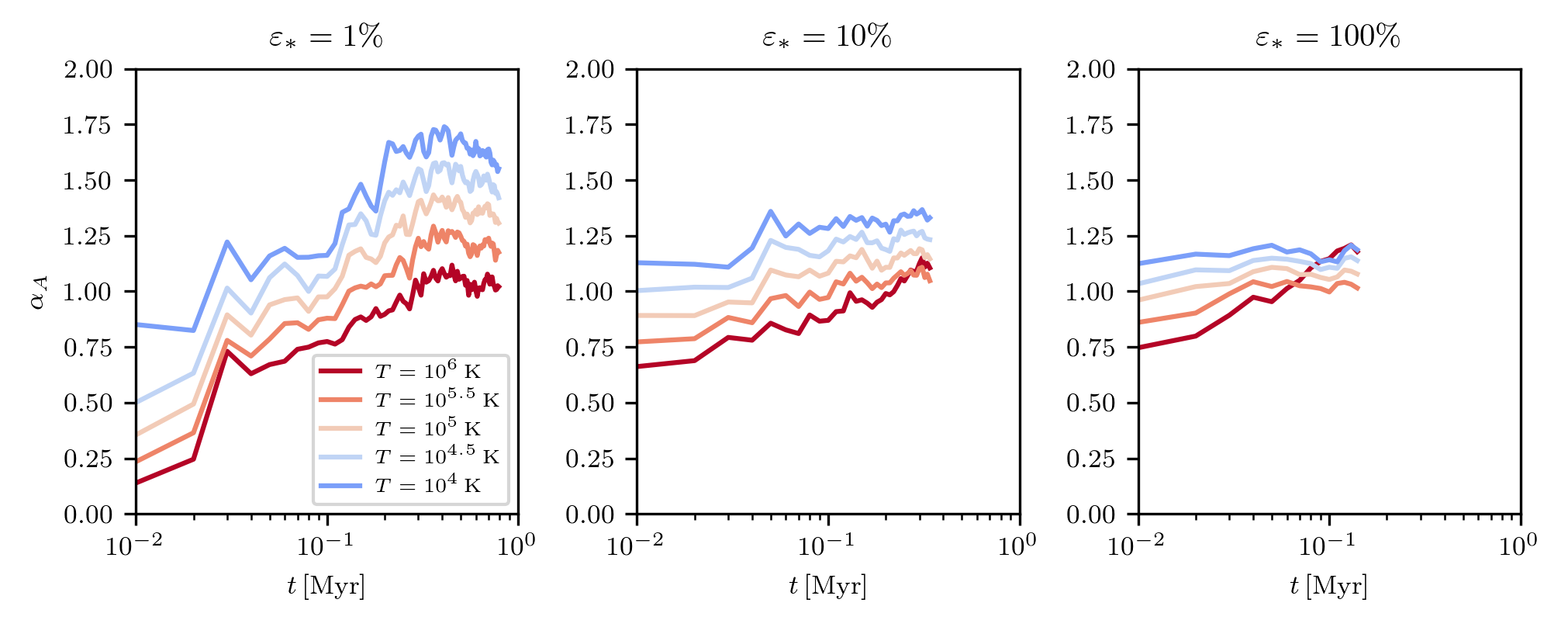}
    \caption{The ratio of bubble surface area $\Abub(T,\ell)$ to fractal 
    area $ 4\pi\reff^2 (\reff/\ell)^{1/2} $ as $\reff(t)$ evolves in time. 
    We only show the evolution up to the point of wind breakout. We set 
    $\ell=32 \delx$ and measure the area for several iso-temperature surfaces 
    as marked in key. Results for $\alpha_A$ from taking the area ratio (see 
    \autoref{eq:areacoeff}) are shown for $\mcloud = 10^5 M_{\odot}$, 
    $\rcloud = 10\, {\rm pc}$ and $\sfe = 1\%, \, 10\%, \,$and$\, 100\%$. The 
    near-constancy of $\alpha_A$ in time, especially at $\sfe = 10\%$ and 
    $100\%$,is an excellent indication that a fractal with $d= 1/2$ describes 
    the geometry of the bubble well.}
    \label{fig:area_alphaA}
\end{figure*}

\subsubsection{Area Measurements}
\label{subsubsec:area_measurement}
Using the results of \autoref{subsubsec:area_fractal}, we can now assess 
the relationship between the bubble's surface area and its effective 
radius, $\reff$, as defined by \autoref{eq:frac_area}. In particular, 
this allows us to test whether a single fractal dimension can describe the 
surface, and to evaluate the free parameter $\alpha_A$ in that equation. 
For this analysis we consider the models with $\rcloud = 10 \, {\rm pc}$ 
and $\mcloud = 10^5 \, M_{\odot}$ at all values of $\sfe$.  

We set 
\begin{equation}\label{eq:areacoeff}
    \alpha_A (\reff, \ell ) = \frac{\Abub(T,\ell)}{4\pi \reff^2 (\reff/\ell)^{1/2}}
\end{equation}
where $\Abub(T,\ell)$ denotes the scale dependent area measurement of the
iso-temperature surface (at a range of $T$), and $\ell$ is the measurement 
scale. We shall use $\ell = 32\delx$, the scale on which the fractal 
structure begins to saturate (as is evident in \autoref{fig:fractal}). 
\autoref{eq:areacoeff} compares the actual area with the prediction for 
a fractal with excess dimension $d=1/2$ as this value is broadly 
consistent across the range of parameters investigated and measurement 
techniques used.

We present the time evolution of \autoref{eq:areacoeff} in 
\autoref{fig:area_alphaA}. Especially for the larger-$\sfe$ cases, we 
conclude that the simple $d = 1/2$ fractal agrees extremely well with 
bubble's true surface area. 

\begin{figure*}
    \centering
    \includegraphics{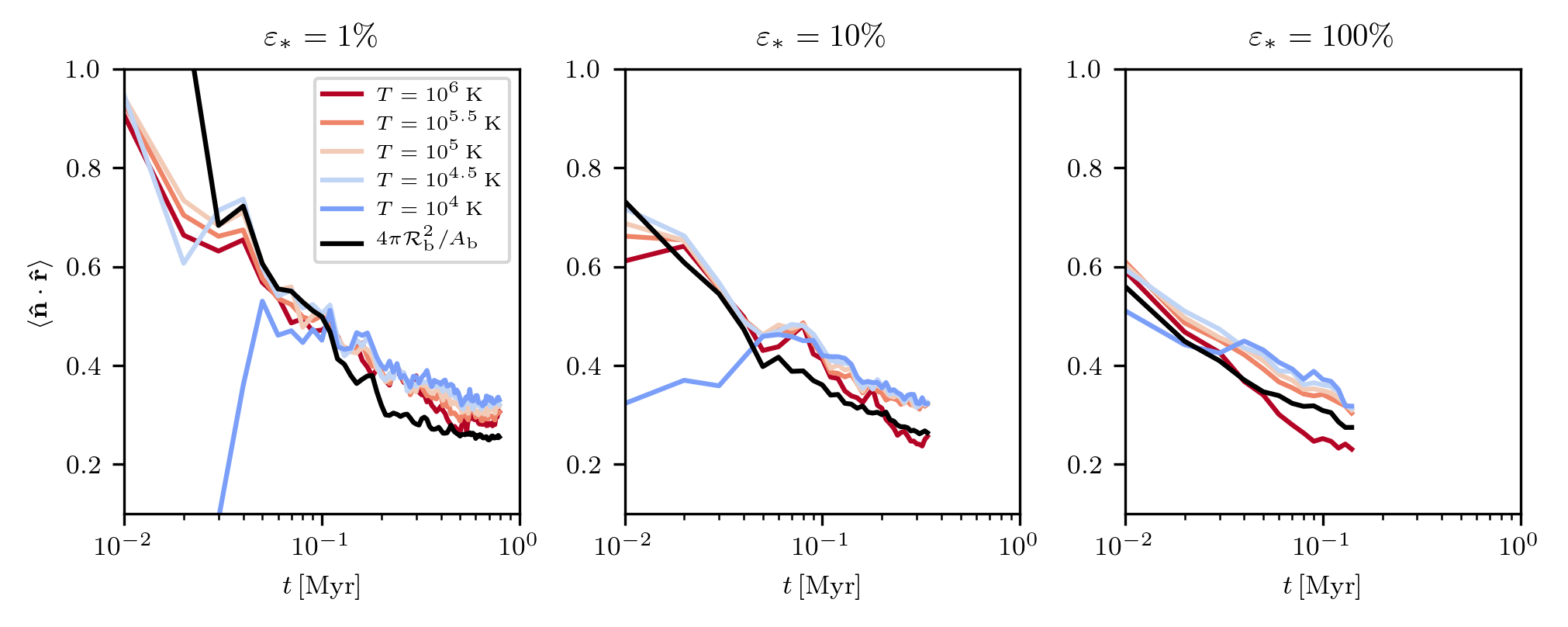}
    \caption{The average value of the dot product of the outward normal 
    vector to the bubble surface with the radial vector, which we term 
    the ``foldedness'' of the bubble surface.  The degree to which this 
    is less than one quantifies the amount that the surface is ``inward 
    facing'' or ``crumpled.'' The same models are used as for 
    \autoref{fig:area_alphaA}, along with the same calculations of the 
    bubble iso-temperature surfaces. We also show 
    $4\pi \reff^2/\Abub(10^6\, {\rm K}, 32\delx)$ in black. The degree to 
    which the black curve matches the other curves reflects how close the 
    effective area, $A_{\rm eff}$ is to the equivalent spherical surface 
    $4\pi \reff^2$. }
    \label{fig:nr_geo}
\end{figure*}

\subsubsection{Bubble Geometry}
\label{subsubsec:bubble_geo}
Finally, we consider an additional characteristic of the bubble geometry, 
which we term its ``foldedness.'' We define this as the value of the dot 
product of the unit normal to the bubble surface ($\nhat$) and the unit 
radial vector ($\rhat$). There are physical reasons that this quantity 
matters. In the spherically symmetric, pressure-driven bubble of 
\citet{Weaver77}, \citet{ElBadry19}, and others, the bubble expands due to 
a pressure gradient force normal to the bubble surface. Since the bubble is 
spherical, the normal vector to the surface is always directed exactly 
radially outwards, so that the pressure force and expansion velocity are 
exactly aligned. In the case of a bubble in a turbulent medium, the surface 
will have a complicated geometry, as \autoref{fig:example_slice_R20} and 
\autoref{fig:example_slice_R2p5} illustrate.  For any pressure force acting 
normal to the bubble surface, only a fraction of the force will act radially 
outward. The radial contribution from a normal force can be then be quantified 
by this foldedness quantity $\foldedness$. 

The quantity $\foldedness$ is our measure of the foldedness of the surface 
in the case of our wind-driven bubbles, and this is important in setting 
the radial force that drives the overall expansion. Specifically, if there 
is a uniform internal pressure $P$, the effective outward radial force on 
the bubble surface can be written as
\begin{equation}
    \label{eq:Fradial}
    F_r \approx P A_{\rm eff}(T,\ell)  \, ,
\end{equation}
for an appropriate effective area, $A_{\rm eff}(T,\ell)$. We define 
$A_{\rm eff}(T,\ell)$ using the area of  iso-temperature surfaces $\Abub(T,\ell)$ 
from \autoref{subsubsec:area_fractal} and the foldedness $\foldedness$ as 
\begin{equation}
    \label{eq:Aeff_def}
    A_{\rm eff}(T,\ell) = \Abub(T,\ell) 
    \foldedness \, .
\end{equation}

Again, using the \texttt{marching-cubes} algorithm, we evaluate $\foldedness$ 
by taking the dot product of the unit normal to each triangular face output by 
the algorithm with the unit radial vector pointing to the center of that 
triangular face. We then average these values over the set of faces. The 
results are displayed in \autoref{fig:nr_geo} for the 
$\mcloud=10^5 M_{\odot}$, $\rcloud=10\,{\rm pc}$ model at all values of 
$\sfe$. In \autoref{fig:nr_geo}, we also show  that the scaling of $\foldedness$ 
in time is very similar to the temporal behaviour of $4\pi \reff^2/\Abub$.
Both of these quantities decrease in time $\propto t^{-1/4} \propto \reff^{-1/2}$.  
But more important than the specific time dependence is the fact that the  scaling of 
the excess fractal dimension is compensated by the foldedness of the surface, such 
that $A_{\rm eff} (T,\ell) \propto \reff^2$.

Our findings on scalings imply that the effective area for radial pressure 
forces is the same as for a spherical surface. We posit that this property is 
generally true. That is, for any bubble, the quantity $A_{\rm eff}$ defined in 
\autoref{eq:Aeff_def} scales as the square of its associated linear scale defined 
through the cube root of volume. 
Though the divergence theorem can be used to show that 
$A_{\rm eff}\equiv \int \hat r \cdot \hat n dA = 4\pi \langle R_{\rm max}^2 \rangle_{\theta,\phi}$ for $R_{\rm max}$ the 
bubble radius at a given spherical polar angle, we have not 
found a proof to connect this to \autoref{eq:reff_def}; nevertheless, it seems intuitively reasonable.

%


The above proposition that $A_{\rm eff} \propto \reff^2$ points to 
the self-consistency of the assumption in the EC solution that the 
angle-averaged radial force in the momentum equation does not 
explicitly depend on detailed geometry, even for a fractal bubble.

It is the combination $\rho v_r^2 + P$ that drives outward expansion 
of the bubble.  The Reynolds stress term is radially directed and 
therefore always contributes to the shell momentum in the $\hat r$ direction, 
but the pressure term would produce a normal force where it acts on the shell, 
and therefore would contribute to the radial momentum as in \autoref{eq:Fradial}.  
Our conclusion that $A_{\rm eff}\propto \reff^2$ is then what allows us to treat 
the total momentum input rate via an equivalent spherical calculation, with 
the breakdown of the two contributing terms provided in the Appendix of Paper I 
(see Equation A11 there).

\begin{figure*}
    \centering
    \includegraphics{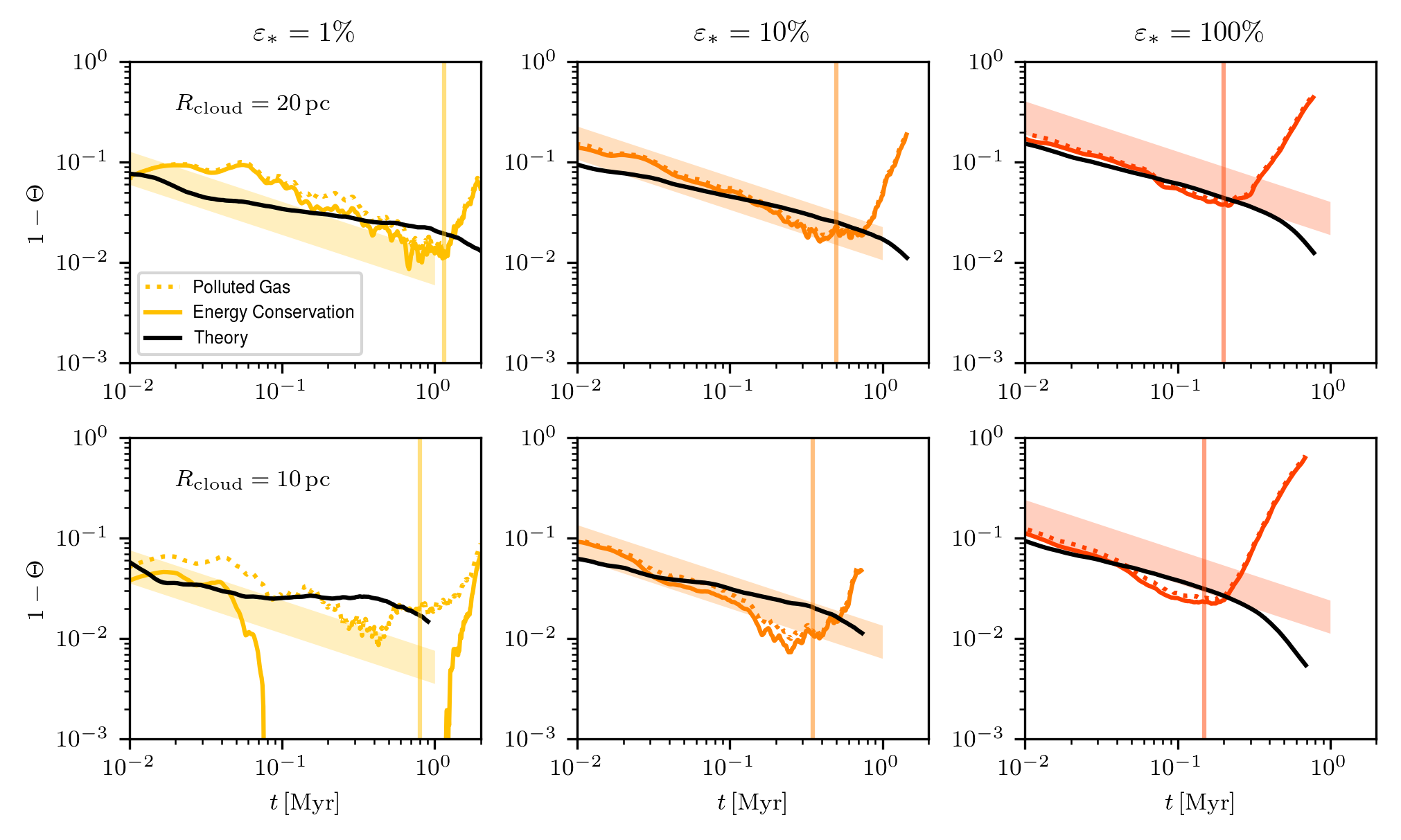}
    \caption{The fraction $1-\Theta$ of the input wind energy that is 
    retained in the bubble and shell, including both kinetic and thermal 
    terms, as a function of time. We show results for the 
    $\rcloud = 20\, {\rm pc}$ in the top row and $\rcloud = 10\, {\rm pc}$ 
    in the bottom row (both with  $\mcloud = 10^5 M_{\odot}$), for  
    $\sfe = 1\%, \, 10\%, \,$and$\, 100\%$. The time when the wind breaks out 
    of the simulation box is denoted in each panel by a colored vertical line. 
    The colored solid lines in each panel denote $1-\Theta$  measured via the 
    energy conservation method, while the dotted lines denote the same quantity 
    measured using cooling in wind-polluted ($\fwind > 10^{-4}$) gas. The shaded areas in each panel 
    delimit the range expected from EC wind theory (see text).  The black line 
    in each panel shows the theoretical prediction from  
    \autoref{eq:Theta_cool_prediction} with values of $f_{\rm turb}$, 
    $\ratr$, $\alpha_p$, and $\alpha_R$ determined as described in \autoref{subsec:params}.}
    \label{fig:cooling}
\end{figure*}

\subsection{Cooling and Energetics}
\label{subsec:energetics}

\subsubsection{Measured Cooling and Retained Energy}
Energy inputs from the wind are split between the (thermal and kinetic)
energy in the interior of the bubble, $\Ebub$, the radial kinetic energy of 
the shell, $\Ersh$, the turbulent kinetic energy in the shell, $E_{\rm turb}$, 
and energy that is lost to cooling, $\ecool$. The EC theory presented in Paper 
I and reviewed in \autoref{sec:theory} provides predictions for $\Ebub$ and 
$\Ersh$ and thus (through the total input energy $\Lwind t$) for the sum of 
the energy lost to cooling and the energy in turbulent motion. The exact split 
between turbulent energy dissipation and cooling is not predicted by our theory, 
but we can account for it in simulations using the quantity $f_{\rm turb}$, 
introduced in \autoref{sec:theory} and shown in \autoref{fig:fturb}. In this 
section we will endeavor to measure the cooling that is occurring in our 
simulations directly and check that this is consistent with the prediction 
given by \autoref{eq:Theta_cool_prediction}.

First, we must measure the cooling directly from the simulations. Since, 
however, there is cooling present in the background gas, mixing between the
background gas and the wind, and cooling occurring in the background gas 
that is shock-accelerated by the wind, this is not a trivial task. The 
measurement is further complicated by the fact that the cooling will be 
a large fraction of $\Lwind$, so that an inclusion of only a small amount 
of background cooling in our measurement can make our inference of 
$\Theta\equiv\dot{E}_{\rm cool}/\Lwind$ greater than 1. To address these 
issues we proceed by measuring the cooling in two separate ways so that we 
may check for consistency between our methods. 

The first way follows our method for measuring 
momentum, as laid out in \autoref{subsubsec:shell_momentum}. Specifically, 
we sum the cooling in all gas with $\fwind > 10^{-4}$ in our simulations. 
The justification of this otherwise arbitrary cut in $\fwind$ is based on 
the excellent agreement among measurements of the radial momentum at 
different resolutions when using this criterion, as evidenced in 
\autoref{fig:pr_comp} and also its agreement with our second measurement 
method, detailed below. 

We measure the cooling in another way by using conservation of energy. The 
time derivative of the total energy in the simulation domain can be written as 
\begin{equation}
    \label{eq:basic_energy_conservation}
    \dot{E} = \Lwind - \ecool - \dot{E}_{\rm out} - \Ebkgnd \, ,
\end{equation}
where $\ecool$ is the sum of cooling that occurs in the swept up ambient gas
and the bubble/shell interface, $\dot{E}_{\rm out}$ is the rate at which energy 
moves out of the simulation domain through the boundaries, and $\Ebkgnd$ is the 
net cooling (= cooling - heating) in the ambient gas that is not due to the 
wind. While it is straightforward to measure the total energy in the simulation 
domain, and hence its time derivative $\dot{E}$, as well as the rate at which 
energy leaves the box, it is not so easy to define what the cooling due to the 
background ambient gas is. 

We address this by running the simulations that we initialized before
turning on a wind (as described in \autoref{sec:run_describe}) forward in 
time without a wind and measuring the net cooling that occurs throughout the 
volume. We will denote the cooling that occurs in these simple turbulent 
evolution simulations as $\dot{E}_{\rm tev}$. To account for the fact that 
part of the simulation volume that would normally be cooling as background 
gas will have been displaced by the wind bubble in our simulations, we 
measure $\Ebkgnd$ as
\begin{equation}
    \Ebkgnd = \left( 1 - \frac{\Vbub}{\Lbox^3}\right) 
    \dot{E}_{\rm tev} \, .
\end{equation}
With this measurement of the background cooling we can use 
\autoref{eq:basic_energy_conservation} to determine $\ecool$.

The results of the two methods for calculating 
$\Theta=\dot{E}_{\rm cool}/\Lwind$ are displayed in \autoref{fig:cooling} 
as solid and dotted lines for the $\mcloud = 10^5 M_{\odot}$, 
$\rcloud = 20$ and  $10\, {\rm pc}$ cases, for all $\sfe$ values.  We only 
show these lower density cases as, in the higher density simulations, the 
background cooling becomes too strong and difficult to separate from the 
cooling associated with the wind, especially at low values of $\sfe$. 
In fact, this effect is already evident in the $\rcloud = 10\, {\rm pc}$, 
$\sfe = 1\%$ case shown in the bottom left panel of \autoref{fig:cooling}.

There is excellent agreement between our two measurement methods, with 
only moderate deviations at low $\sfe$ values. The dramatic increase in 
$1-\Theta$ (decrease in cooling) towards the end of each simulation can 
be attributed to breakout and venting of the wind outside of the simulation 
domain, the onset of which is indicated by the colored vertical line.

In \autoref{fig:cooling} we also show, as indicated by the shaded region in 
each panel, where we expect the solutions for $1- \Theta$ to lie. This is 
bounded above by the locus at which the EC condition applies as given by 
\autoref{eq:cooling_condition}, and bounded below  the 
locus of maximum cooling as given by \autoref{eq:Theta_cool_prediction} with 
$f_{\rm turb} = 0$ and $\ratr = \alpha_R =\alpha_p =  1$, so that the 
right-hand side becomes $(3/2)\dot {R}_{\rm EC}/\vw$. Both of these limits 
have retained energy fraction $1-\Theta \propto \dot R_{\rm EC}/\vw $ with 
different coefficients (3.8 and 1.5, respectively), decreasing in time as 
$1-\Theta \propto t^{-1/2}$. Finally, as black lines, we show the prediction 
of \autoref{eq:Theta_cool_prediction} in its entirety, with time-varying 
values for $f_{\rm turb}$, $\ratr$, $\alpha_p$, and $\alpha_R$ measured 
as described in \autoref{subsec:params}.  

Overall, \autoref{fig:cooling} shows that the different numerical 
measurements of $1-\Theta$ are in good agreement with each other and with
theoretical estimates. At early times ($\sim 10^4\yr$), the energy retained 
in the bubble amounts to $\sim 10\%$ of the input, slightly increasing for 
lower density clouds and for more powerful winds (larger $\sfe$).  This 
decreases $\propto t^{-1/2}$ until the time breakout occurs, 
$t\sim 10^5-10^6\yr$ (decreasing at higher wind power and for denser clouds).  

\begin{figure}
    \centering
    \includegraphics{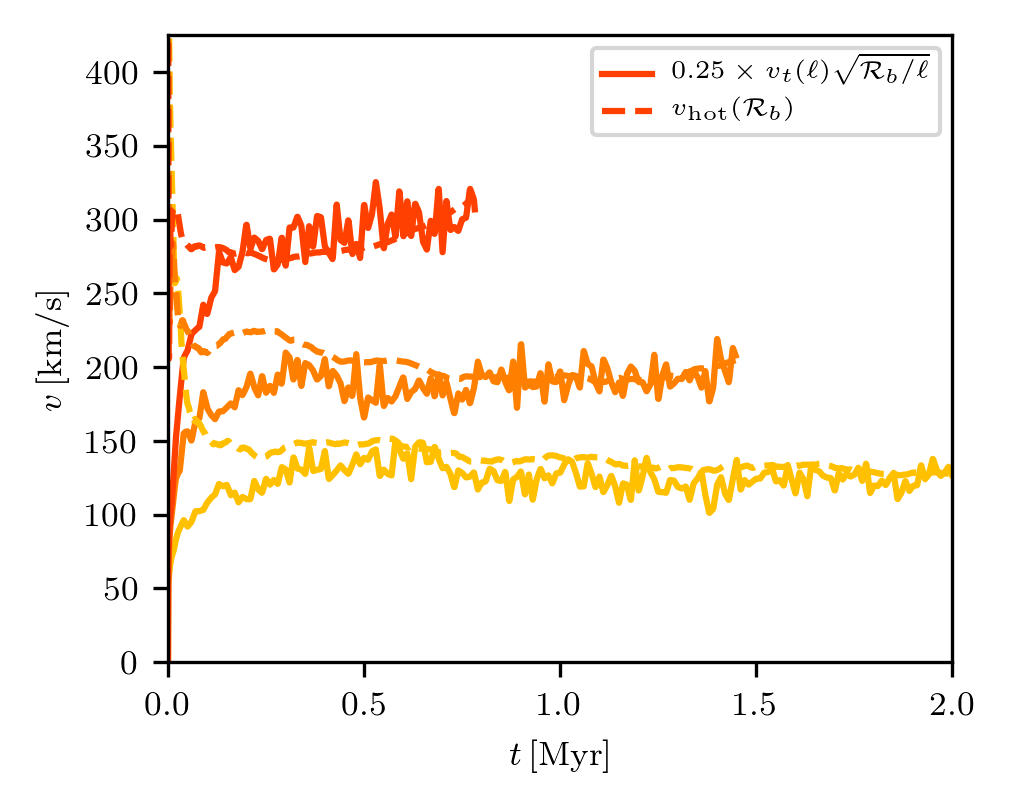}
    \caption{Evolution of quantities related to the flow into and through 
    the turbulent mixing/cooling boundary layer between the bubble and the 
    background. The expected radial velocity (\autoref{eq:vrb}) of shocked wind 
    gas advected into the fractal interface is shown as dashed lines. The 
    ``equivalent velocity'' of gas transiting the mixing/cooling layer 
    (\autoref{eq:veff} with $d=1/2$) is shown as solid lines, including a factor 
    of $0.25$ chosen to match the dashed lines. These curves are shown for the 
    model with $\mcloud = 10^5 M_{\odot}$, $\rcloud = 20$, and all $\sfe$. We 
    use $\ell_{\rm min} = 8 \delx$ for the measurement scale of the turbulent 
    structure function.}
    \label{fig:vt_capacity}
\end{figure}

\subsubsection{Total Possible Cooling}
\label{subsubsec:fractal_cooling_estimate}

Finally, we connect our results to the expected requirements for the EC 
solution based on the theory of cooling and mixing at turbulent, fractal 
interfaces, as discussed in Paper I. To this end, we measure two 
quantities.  The first is the expected radial velocity of shocked wind gas 
being advected to the boundary mixing layer ($v_{\rm hot}(\reff)$, given by 
\autoref{eq:vrb}) The second is the ``equivalent velocity'' of gas flowing 
through the mixing/cooling layer, as given by \autoref{eq:veff}; this takes 
into account the fractal structure of the bubble surface, but is only 
really defined up to a multiplicative factor from our theoretical analysis.

For this exercise, we use the time dependent measurements for the evolution 
of the bubble's effective radius (\autoref{fig:radius_comp}), the shock radius 
($\rfree$), and the turbulent structure function (\autoref{fig:vt_struct_ev}) 
in models with $\mcloud = 10^5 M_{\odot}$, $\rcloud = 20$, and all $\sfe$.  We 
choose $\ell_{\rm min} = 8 \delx$, and calculate 
$v_{\rm equiv}=\vt (\ell_{\rm min}) ({\reff/\ell_{\rm min}})^{1/2}$. Note that we 
have implicitly adopted a fractal dimension of $d=1/2$, which is broadly consistent 
with our results in \autoref{subsubsec:cooling_fractal} and 
\autoref{subsubsec:area_fractal}. As noted above, the ``equivalent velocity'' 
in \autoref{eq:veff} is only predicted up to an order-unity coefficient.  We 
therefore multiply $v_{\rm equiv}$ by a factor $0.25$, and show the results of 
this calculation in \autoref{fig:vt_capacity}.

\autoref{fig:vt_capacity} shows that there is excellent agreement between 
the expected velocity at which energy and mass arrives at the surface of the 
bubble (an advection speed), and the effective flow velocity through the 
mixing/cooling boundary layer (a diffusion speed). This directly demonstrates a 
key feature of our model: all thermal energy that is delivered to the turbulent 
interface is efficiently mixed in and radiated away as rapidly as it arrives.

\section{Summary and Conclusion}
\label{sec:conclusion}

In Paper I we described a theory for the expansion of a stellar wind-driven 
bubble into the dense, turbulent ISM, characterized by strong cooling losses 
due to turbulent mixing of the hot gas with denser gas at the bubble surface. 
We posited that this cooling was large enough to cause the dominant phase of 
the bubble's evolution to be momentum driven.

A solution in which momentum input (rather than energy input) controls bubble 
evolution has been discussed in the past by several authors in various contexts 
\citep{Steigman75,OstrikerMcKee88,KooMcKee92a,KooMcKee92b,KimOstrikerRaileanu17,SilichTT13}, 
and others have suggested that such solutions could be associated with efficient 
mixing and cooling at boundary layers 
\citep{GarciaSegura96,DaleBonnell08,Mackey15,Fierlinger16}, but none have previously 
demonstrated that this regime generally applies within star-forming molecular 
clouds. We do this by conducting a large suite of 3D hydrodynamic simulations 
with winds injected into dense, turbulent ISM material. Analysis of our 
simulations shows that the predictions of our theory very accurately describe 
the evolution of the wind bubble's volume, the momentum that it carries, and its 
energetics. Our simulations demonstrate that the limit of maximally efficient 
cooling in our theory is most appropriate for the strongest stellar winds 
(parameterized in our model by high values of the star formation efficiency $\sfe$).

Our theory and simulations explore in detail where energy is stored and explain 
physically how most of it is radiated away, via processes analogous to those that 
have been investigated in recent simulations of Kelvin-Helmholtz unstable mixing 
layers with fractal geometries \citep[e.g.][]{FieldingFractal20,Tan20}.  The fractal 
theory of the bubble interface also accurately predicts essential quantities of 
the bubble such as the area of its interface with the ambient ISM and properties 
of the geometry of this interface. We additionally demonstrate that it is the 
fractal nature of the bubble that allows for such efficient cooling by showing 
that the cooling capacity under our fractal theory is greater than the energy 
being provided by the stellar wind.

The main conclusions from our simulations can be summarized as follows:
\begin{enumerate}
    \item The effective or mean bubble radius expands in a 
    momentum-driven manner as $\reff \propto t^{1/2}$ 
    (\autoref{eq:rbub_ecw}). This is shallower than the original 
    \citet{Weaver77} solution or the modified \citet{ElBadry19} 
    solution ($\reff \propto t^{3/5}$).
    
    \item  The total momentum carried by the bubble increases approximately 
    linearly in time (\autoref{eq:pEC}) and is typically only slightly larger 
    (at most a factor 4) than the total momentum injected by the wind, with the 
    smallest enhancement for the most luminous winds (largest $\sfe$), as 
    shown in \autoref{fig:pr_comp}. The momentum enhancement factor $\alpha_p$ 
    reflects buildup of hot shocked gas within the bubble; higher turbulence levels 
    in models with more powerful winds drive stronger interface mixing and limit 
    this buildup, keeping $\alpha_p$ very close to unity.
    
    \item  From analysis of the shells in our simulations, we find (see 
    \autoref{fig:fturb}) that the energy carried in tangential motion is 
    typically greater than that carried in radial motion (turbulence dominated). 
    These results show that shells typically become more turbulence-dominated in 
    time up to the point of bubble `breakout' from the simulation domain. We also 
    find that cases with less luminous winds and denser environments are more 
    turbulence dominated.
    
    \item  We measure the turbulence in the hot gas directly, and find typical 
    velocities of $150-350\, {\rm km/s}$ (amplitude increasing with input momentum 
    rate; see \autoref{fig:structure_functions}), which are about $10-15\%$ of the 
    wind velocity $\vw$. This turbulence has characteristic energy-containing or 
    outer scale that is larger in larger clouds (\autoref{fig:structure_functions}), 
    suggesting that the background inhomogeneity induced by turbulence helps to 
    define it.  The outer scale is typically $\sim 10\%$ of the bubble size, 
    growing with the bubble (\autoref{fig:vt_struct_ev}).  

    \item  We quantify the excess fractal dimension $d$ of the shell-bubble 
    interface in two different ways (see \autoref{fig:fractal}), and find that 
    this generally falls in the range of $d= 0.4-0.7$. Lower luminosity winds 
    generally have higher fractal dimensions, which is consistent with the 
    shells driven by these winds being more turbulence dominated, as it is the 
    turbulence that induces the fractal structure. We also find that the fractal 
    characterization of the shell interface is in excellent agreement with the 
    measured area (\autoref{fig:area_alphaA}) and geometry (\autoref{fig:nr_geo}).
    
    \item  We use $1-\Theta$ to denote the fraction of the energy input rate
    from the wind that remains in the bubble as either thermal or kinetic energy at 
    any given time. We find that $1-\Theta \sim 0.1 -0.01$, decreasing in time  
    $\propto t^{-1/2}$ (\autoref{eq:Theta_cool_prediction}) for all models 
    (\autoref{fig:cooling}).  The very small value of $1-\Theta$ is consistent with observational constraints 
    (see Paper I discussion).
\end{enumerate}

We have demonstrated that our simulations are well converged in all of the 
quantities that we investigate here.  Higher resolution should only increase cooling, 
so we believe our results for the efficient-cooling bubble evolution solution are 
robust. However, our present simulations do not resolve the scale $\lcool$ at the 
wind bubble interface where mixing and cooling timescales would be comparable in the 
real ISM. Full validation of the small scale processes discussed here will therefore 
require higher resolution simulations, an important direction for future work.

Finally, we reiterate that the simulations presented here include only a limited set of the relevant physical processes.  In particular, we have not included radiation  or magnetic fields.  EUV radiation would photoevaporate gas from the shell surfaces that ``face'' the cluster, while  FUV radiation absorbed by dust (where it is not destroyed) directly deposits photon momentum; both effects drive expansion of the bubble  surrounding the cluster.  The free wind and shocked wind regions would then in  general interact with photoionized gas that has been evaporated from the shell, rather than directly with the  denser shell gas \citep[e.g.][]{Dwarkadas13,Geen20}.  With a lower density contrast, Kelvin-Helmholz instabilities at hot/warm interfaces would have higher growth rates than for hot/cold interfaces at the same pressure.  The turbulent mixing process would likely occupy a larger volume of photoionized gas compared to the situation here, since the turbulence levels in the interaction region depend on the density contrast \citep{FieldingFractal20}.    Magnetic fields presumably also affect the mixing/cooling process.  Given the extremely high velocities of the shocked wind, field strengths in the cloud would have to be very (unrealistically) high to prevent primary  instabilities at the interface, but magnetic tension could still limit the turbulent cascade at small scales (and render it anisotropic). Numerical magnetohydrodynamic studies including both winds and radiation will be needed to assess how the turbulent mixing/cooling process and overall dynamical evolution are quantitatively affected.     

\acknowledgments

We thank Drummond Fielding and Erin Kado-Fong for useful discussions. 
We thank the referee for their many insightful comments that improved 
the quality of the manuscript. This work was partly supported by the 
National Science Foundation (AARG award AST-1713949) and NASA (ATP 
grant No. NNX17AG26G). J.-G.K. acknowledges support from the Lyman 
Spitzer, Jr. Postdoctoral Fellowship at Princeton University. 
Computational resources were provided by the Princeton Institute for 
Computational Science and Engineering (PICSciE) and the Office of Information 
Technology’s High Performance Computing Center at Princeton University.

\software{
{\tt Athena} \citep{Stone08_Athena,StoneGardiner09},
{\tt astropy} \citep{astropy13,astropy18}, 
{\tt scipy} \citep{scipy},
{\tt numpy} \citep{harrisNumpy2020}, 
{\tt IPython} \citep{Perez07}, 
{\tt matplotlib} \citep{matplotlib_hunter07},
{\tt xarray} \citep{hoyer17}, 
{\tt pandas} \citep{pandas2020},
{\tt adstex} (\url{https://github.com/yymao/adstex})
}

\clearpage
\appendix

\section{Validation of Wind Implementation}
\label{app:validation}

\begin{figure*}
    \centering
    \includegraphics{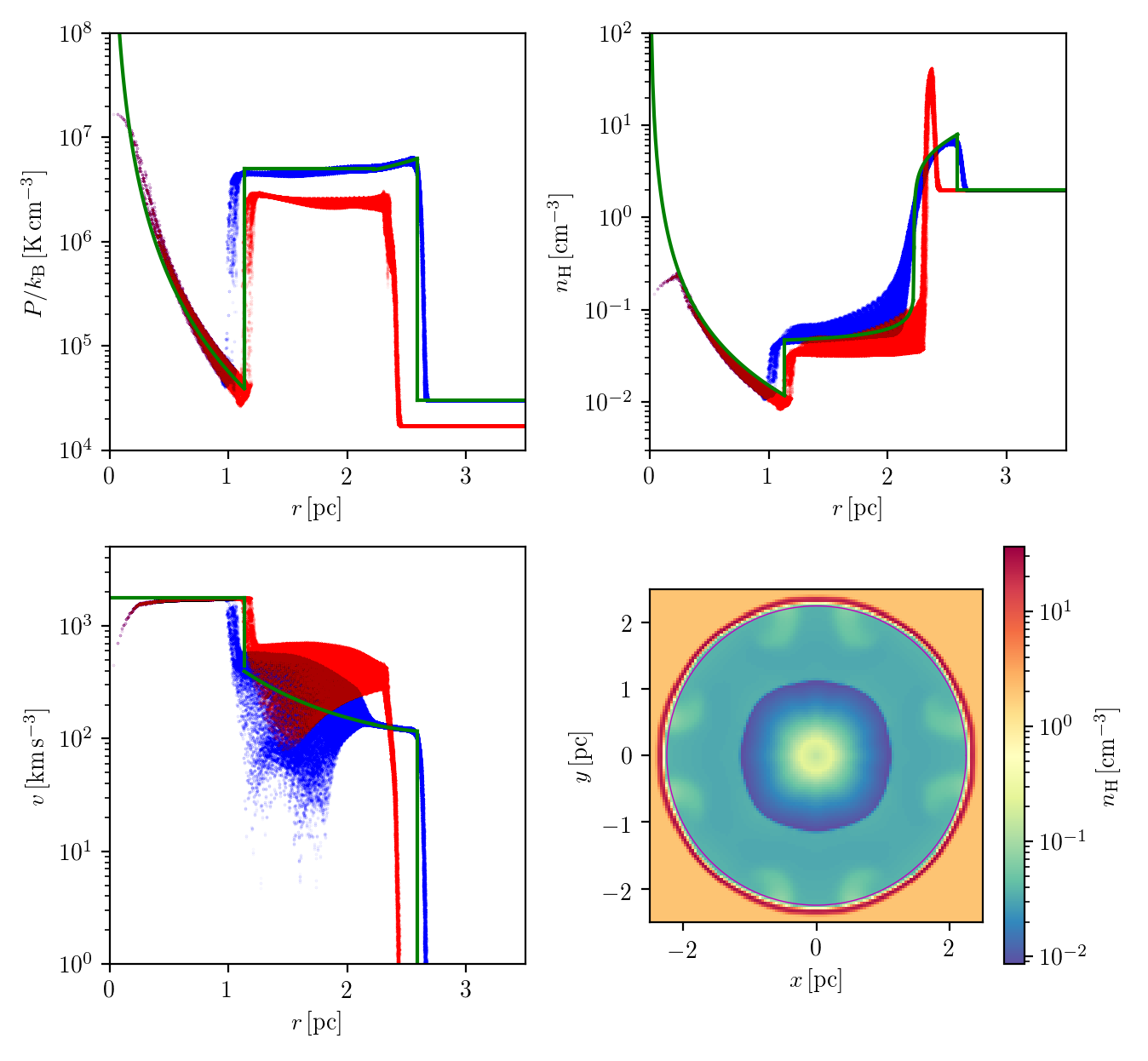}
    \caption{Wind implementation tests
    for a uniform background, shown for a  snapshot $10^4\, {\rm yrs}$ 
    after the start of wind feedback. We show profiles (every cell
    is plotted) of pressure (top left), number density (top right), 
    and velocity magnitude (bottom left) in radius for both the 
    adiabatic run (in blue) and the run with cooling (red). The 
    \citet{Weaver77} solution for fully adiabatic evolution is shown in 
    green for these same panels. In the bottom right panel we show a 
    slice of number density through the $z=0$ plane of the simulation 
    for the run with cooling. One can clearly see the cooled, dense 
    shell that has formed at the outer edge of the wind bubble. The 
    \citet{Weaver77} solution for the radius of the shell in the case 
    that gas in the leading shock cools is 
    indicated as a magenta circle.}
    \label{fig:windtest}
\end{figure*}

We test our implementation of stellar winds by comparing to 
analytic solutions given by \citet{Weaver77} for a spherical wind expanding 
into a uniform, static background medium. In our test the background medium 
has a number density of $n_{\rm H} = 2\, {\rm cm}^{-3}$. The wind has a 
mechanical luminosity of $\Lwind = 10^{37} \, {\rm ergs} \, {\rm s}^{-1}$ 
and a mass loss rate of $\dot{M}_w = 10^{-5} M_{\odot} \, {\rm yr}^{-1}$.
This corresponds to a wind velocity of 
$\vw = 1.78\times 10^3 \,{\rm km}\,{\rm s}^{-1}$.
The simulation was run in a $L = 5\,{\rm pc}$ box with $128^3$ cells. 
The feedback radius was set as $\rfb = 0.25 \,{\rm pc}$ 
($\rfb/\delx = 6.4$), and we used $\Nsc=4$ subcells. 
We run simulations both with and without cooling (the latter is fully 
adiabatic).  The results of these tests are 
displayed in \autoref{fig:windtest}. 

Comparison of the profile for the \citet{Weaver77} fully adiabatic 
solution (described in Section 2 of that paper, shown in green) and 
our adiabatic simulation (shown in blue) makes clear that the code 
accurately captures the expected behaviour. In the realistic 
non-adiabatic case the shell of swept up ambient gas is expected to 
cool and collapse once the cooling time in the shell is much shorter 
than the lifetime of the wind. The profile of the simulation with 
cooling (in red) shows that this has occurred, and that the dense 
shell is at smaller radius than in the adiabatic case. From the density 
slice of the non-adiabatic simulation shown in the bottom right panel of 
\autoref{fig:windtest}, it is clear that this solution matches the expected 
evolution of the non-adiabatic phase of wind evolution (where the expected 
radius of the shell in this case, given by Equation 1 of Paper I, is shown 
as a magenta circle) and retains a high degree of spherical symmetry.

\section{Additional Tests}
\label{app:tests}
In this appendix we display various tests and extensions of 
our theory.

\subsection{Pure Thermal Feedback}
\label{app:thermal_feedback}

As described in \autoref{sec:methods}, we use a hybrid thermal/kinetic 
energy injection scheme which provides for larger time-steps while at 
the same time avoiding issues of discontinuity in the injected energy 
field near the source particle. However, in a real star cluster, winds 
initially emanate from a few massive stars in the form of bulk kinetic 
energy through line driving \citep{LucySolomon70,Abbott82,Sundqvist14}.
The winds from these separate stars then collide and shock-heat,
converting the initial kinetic energy into thermal energy within the
cluster region \citep{Krause13,Canto00}. The concentrated thermal 
energy is then yet again converted into bulk kinetic energy as pressure 
gradients produce acceleration on slightly larger scale, as in 
\citep{ChevalierClegg85}. Since the energy injection zones in our 
simulations are comparable to the size of a star cluster, perhaps 
a more realistic scheme would be simply to inject purely thermal 
energy into the feedback region. This is also interesting to test 
given that thermal energy is lost very  efficiently in our model 
(but only at the outer edge of the bubble). 

\begin{figure*}
    \centering
    \includegraphics{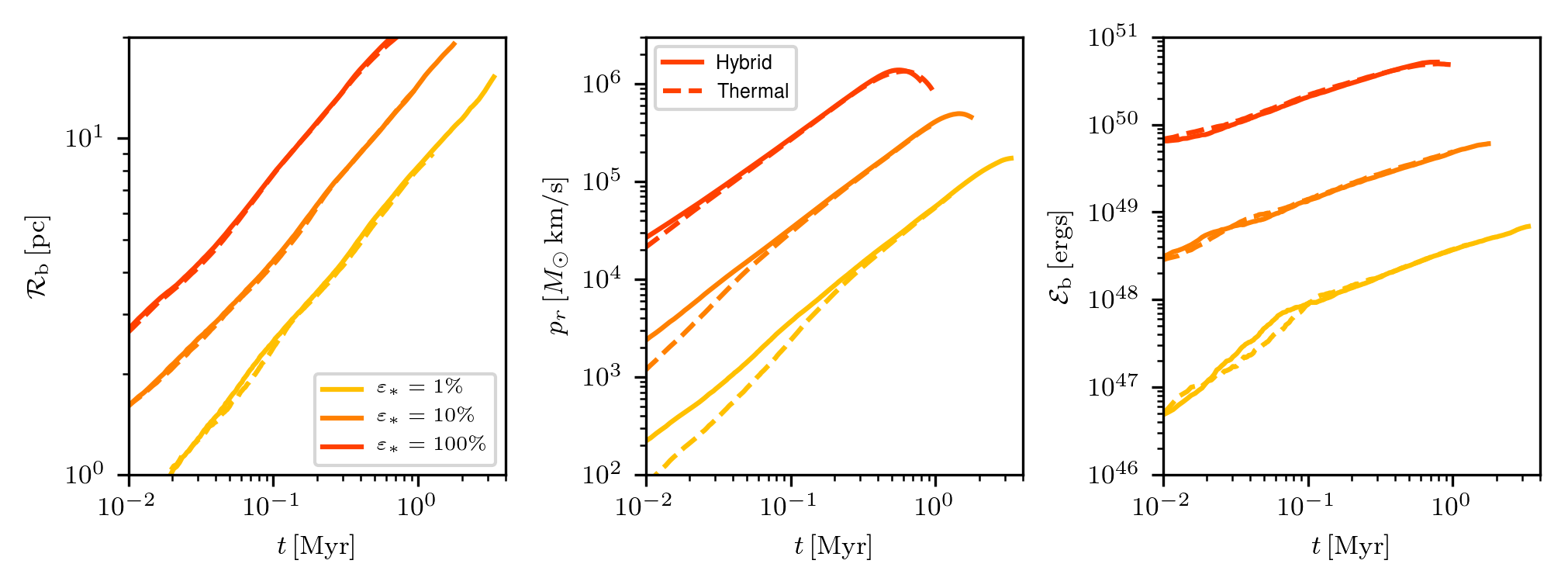}
    \caption{Tests run with purely 
    thermal energy injection, in comparison to the hybrid scheme 
    explained in the main text. All panels are analogous to the 
    top panels shown in \autoref{fig:reduced_mdot_test}. Results  
    using the hybrid injection scheme (thermal scheme) are shown 
    with solid (dashed) lines. The evolution of the total radial 
    momentum, $\pr$ is very slightly delayed for the thermal scheme 
    compared to the hybrid feedback mechanism, but reaches the same 
    scaling by the time the bubble radius is $2-3\pc$.}
    \label{fig:therm_comp}
\end{figure*}

To this end we perform the same simulations described in 
Appendix~\ref{app:reduced_mass_loss} except adopting the value of $\mdot/\ms$ 
used in the main text, and now injecting purely thermal energy rather 
than using the hybrid injection scheme described in \autoref{sec:methods}. 
The results of these tests are shown in \autoref{fig:therm_comp} for the 
main parameters of interest ($\reff$, $\pr$, and $\Ebub$), in comparison 
with results for the hybrid injection scheme. The simulations with thermal 
energy injection lag those with the hybrid energy injection, but only very 
slightly. This is most noticeable for radial momentum, $\pr$, while $\reff$ 
and $\Ebub$ are both relatively insensitive to this change. These results 
indicate that the bubble evolution is not strongly dependent on the energy 
injection method.

\subsection{Reduced Mass Loss}
\label{app:reduced_mass_loss}

For practical reasons, our simulations use a value for the 
wind mass loss rate per unit stellar mass, $\mdot/\ms$, that 
is about a factor of 5 larger than that predicted by the 
SB99 code for a Kroupa IMF. From 
\autoref{fig:sb99_comp}, this results in a wind momentum 
injection rate a factor of $\sqrt{5}\approx 2.2$ larger than 
from SB99. To confirm that our theory still applies when 
using a lower value for $\mdot/\ms$, we ran tests with 
$\mdot/\ms = 2\times 10^{-3} \, {\rm Myr}^{-1}$, similar to 
the SB99 level. This correspondingly reduces the momentum 
injection rate to 
$\pdot/M_* = 7.96 \, {\rm km} \, {\rm s}^{-1} \, {\rm Myr}^{-1}$.  
With a lower momentum injection rate, the right-hand side of 
\autoref{eq:cooling_condition} is reduced, and more cooling 
(larger $\Theta$) would be required to satisfy the condition 
for efficient cooling.  If this EC condition is not satisfied, 
energy will build within the bubble, so that the EC predictions 
for radius, momentum, and energy would underestimate the true values.

For the reduced mass-loss rate, we run three simulations ($\sfe=1$, $10$, $100\%$)
with $\rcloud = 20\, {\rm pc}$, $\mcloud = 10^5 M_{\odot}$, and $\Lbox/\delx = 128$. 
We note that the choice $\rcloud = 20\, {\rm pc}$ (the low density regime) is the 
\textit{worst case scenario} for application of the EC theory since cooling is least 
efficient at low densities, already making it harder for 
\autoref{eq:cooling_condition} to be satisfied.

\begin{figure*}
    \centering
    \includegraphics{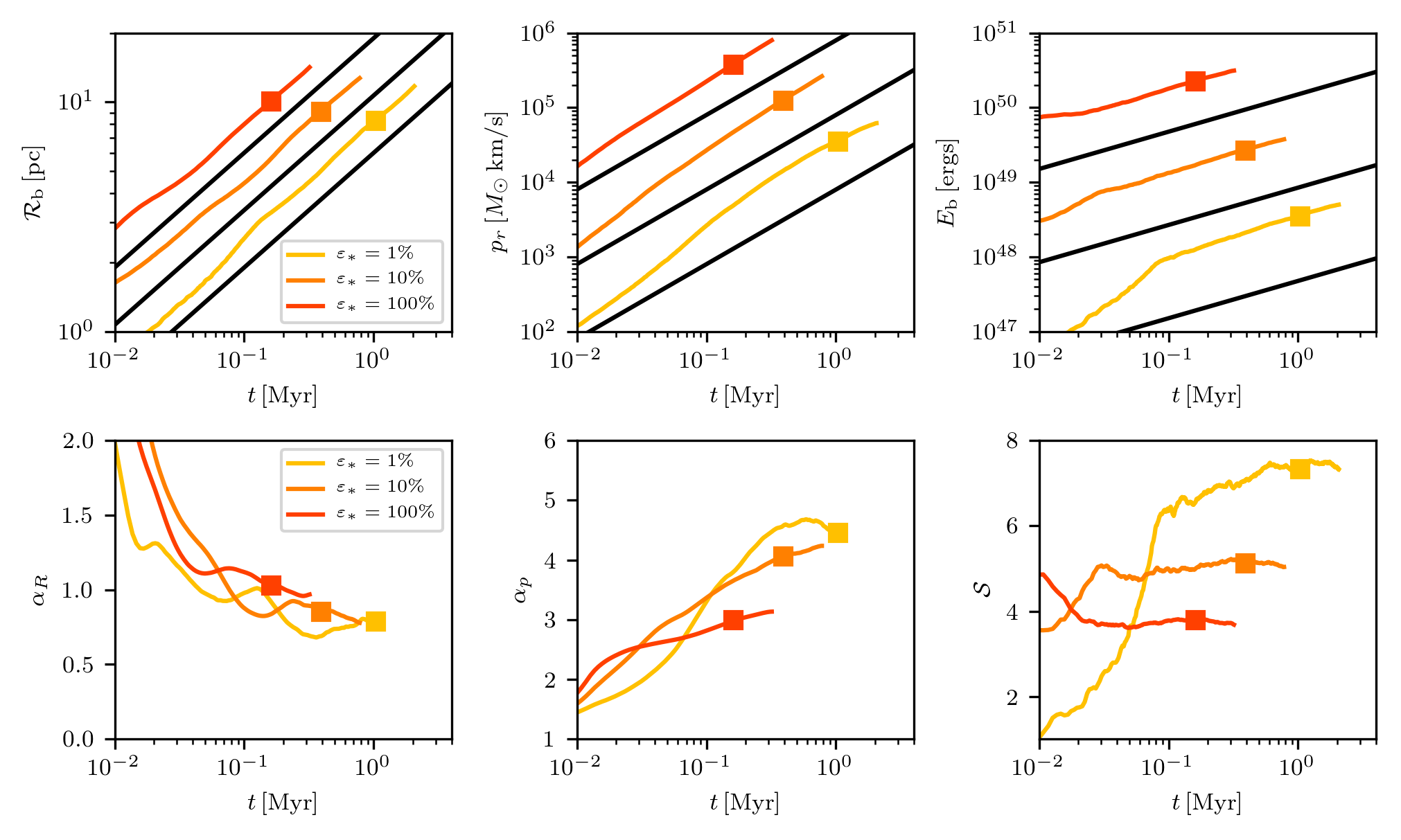}
    \caption{Results of our test runs with smaller mass loss 
    rates ($\mdot/\ms$).  The top panels show comparisons to the EC 
    theoretical predictions (black) for the bubble's effective radius (left), 
    total radial momentum (middle), and total energy in the bubble interior 
    (right) for simulations run with $\sfe = 1\%$ (yellow), $10\%$ (orange), 
    and $100\%$ (red). The bottom panels show quantities related to the ratio 
    of the simulated and theoretically predicted values 
    (see Section~\ref{subsec:params}).}
    \label{fig:reduced_mdot_test}
\end{figure*}

The evolution of the bubble's effective radius ($\reff$), the total momentum 
carried by the bubble ($\pr$), and the total bubble internal energy ($\Ebub$) 
are shown in the top panels of \autoref{fig:reduced_mdot_test}. Similar to the 
figures shown in \autoref{sec:results}, we display the quantities derived from 
simulations using yellow, orange, and red lines compared to the theoretical 
predictions of the EC model in black. As expected, the EC theory underestimates 
$\reff$, $\pr$, and $\Ebub$. However, these differences are small, especially 
in the case of the radial evolution. 

Moreover, the temporal dependence of each quantity is still very well 
described by the EC theory. This is exemplified by the bottom panels of 
\autoref{fig:reduced_mdot_test}, which display quantities related to 
the ratio of the simulated values to the values predicted by our EC theory.
The fact that these quantities are roughly constant and order-unity 
over the course of the simulation, especially for higher-luminosity clusters, reflects that the EC theory still describes evolution reasonably well.

\subsection{Changed Feedback Radius}
\label{app:rfb_test}

For all the simulations in the main text, we keep the feedback radius, 
which determines the region where wind energy is injected, 
constant. As a further test of numerical robustness and convergence, 
we perform a set of simulations with a feedback radius that is half that 
given in the main text, $\rfb = 0.5 \, {\rm pc}$. The set of models is as 
described in Appendix~\ref{app:reduced_mass_loss} but with the normal 
$\mdot/\ms$ values described in the text and now with 
$\Lbox/\delx = 256$, so that $\rfb/\delx \gtrsim 2$ (i.e. the feedback region is still 
well resolved).  We compare to the standard simulations with 
$\rfb = 1 \, {\rm pc}$ and $\Lbox/\delx = 256$.

\begin{figure*}
    \centering
    \includegraphics{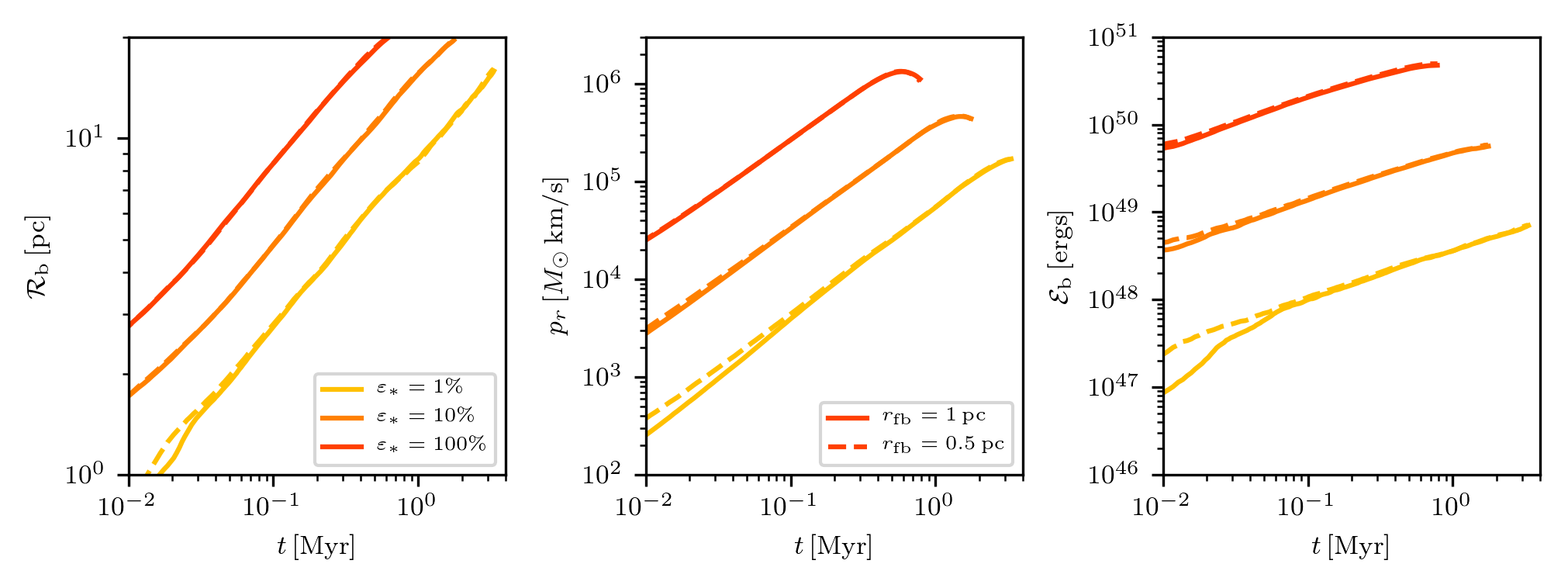}
    \caption{Results of tests with standard ($\rfb = 1 \, {\rm pc}$, 
    solid curve) and and smaller ($\rfb = 0.5 \, {\rm pc}$, dashed curve) 
    feedback radius.  All panels are analogous to those shown in 
    \autoref{fig:therm_comp}.  There is excellent agreement amongst all 
    quantities, indicating that the changed feedback radius (higher source
    resolution) does not impact our results.}
    \label{fig:halfrf_comp}
\end{figure*}

The results of the comparison between $\rfb = 1 \, {\rm pc}$ and $\rfb = 0.5 \, {\rm pc}$ cases is shown in \autoref{fig:halfrf_comp} 
for the standard quantities of relevance to our theory. As expected, 
the results for the smaller feedback radius simulations appear to enter 
the scaling regime earlier than those simulations with larger $\rfb$.

\subsection{Different Turbulent Initial Conditions}
\label{app:turbIC_test}

When initializing the velocity fields in our simulations, we generally use 
the same set of amplitudes and phases (based on the same sequence of random 
seeds) to create the turbulent velocity field. With a different initial 
velocity field (even for the same initial kinetic energy), the background 
density structure into which the bubble expands would be different. To check 
whether the results are sensitive to the specific realization of the turbulent 
velocity field, we perform the same simulations as in 
Appendix~\ref{app:reduced_mass_loss}, using the standard $\mdot/\ms$, except we 
initialize the velocity field with a different random seed. 

While the cloud will still have the same statistical density structure, the 
gas in the immediate vicinity of the star particle will likely be different. 
We expect this to result in a slightly different early-time evolution of the 
wind bubble, while the same overall evolution would be followed once the bubble 
has probed a significant fraction of the cloud size. In reality, the massive 
stars that create these high-powered winds will preferentially form at density 
maxima, but such a self-consistent treatment is left for later work. 

\begin{figure*}
    \centering
    \includegraphics{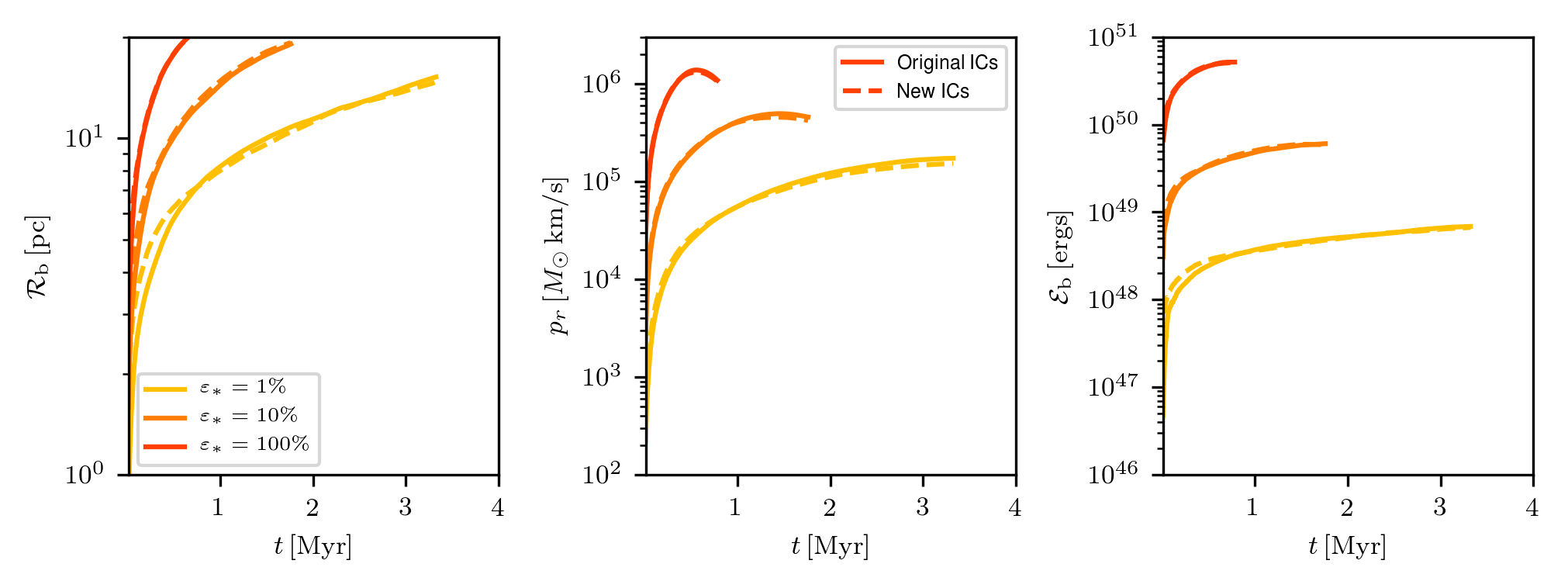}
    \caption{Results of our test runs comparing different initial 
    realizations of the  turbulent velocity field, which creates 
    different cloud density structure. All panels are analogous 
    to those shown in \autoref{fig:therm_comp} except here we use 
    a linear scale for time.  This choice is so as to not overemphasize 
    differences in the early evolution which are due to differences in the 
    density structure very near to the source particle. We show the original 
    initialization as solid lines and the alternate initialization as 
    dashed lines. There is clearly very good agreement at later times between 
    the evolution, in spite of different density structures.}
    \label{fig:turbIC_comp}
\end{figure*}

The results of these tests are shown in \autoref{fig:turbIC_comp}. So as 
not to over-emphasize the differences at early times, we display these results 
on a linear (rather than logarithmic) time scale. Comparing to the standard 
simulation with a different set of turbulent phases, we see that indeed the 
overall evolution of the wind-driven bubbles is insensitive to the specific 
density structure within the turbulent cloud.

\begin{figure*}
    \centering
    \includegraphics{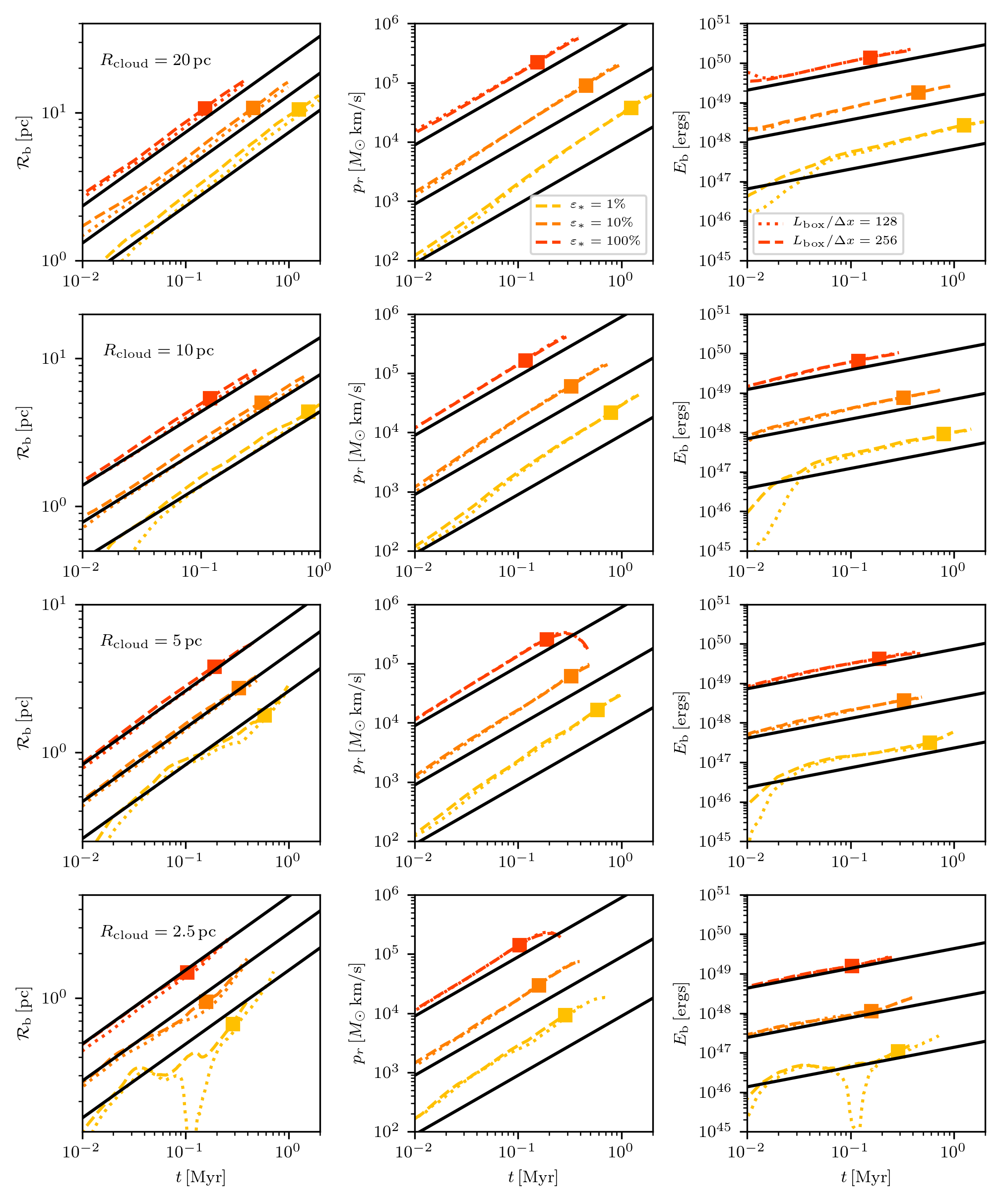}
    \caption{Comparisons of our theory (black curves) to 
    our simulations (colored curves) for cases with cloud mass 
    $\mcloud = 5\times 10^4 M_{\odot}$ and (from top to bottom row) 
    $\rcloud = 20 \, {\rm pc}$, $10 \, {\rm pc}$, $5 \, {\rm pc}$, 
    and  $2.5 \, {\rm pc}$. Columns from left to right show the 
    bubble's effective radius, $\reff$, the total radial 
    momentum, $\pr$,  and the total energy in the 
    bubble interior $\Ebub$. All curves are calculated 
    exactly analogously to those in \autoref{fig:radius_comp}, 
    \autoref{fig:pr_comp}, and \autoref{fig:energy_comp}.}
    \label{fig:m5e4_summary}
\end{figure*}

\begin{figure*}
    \centering
    \includegraphics{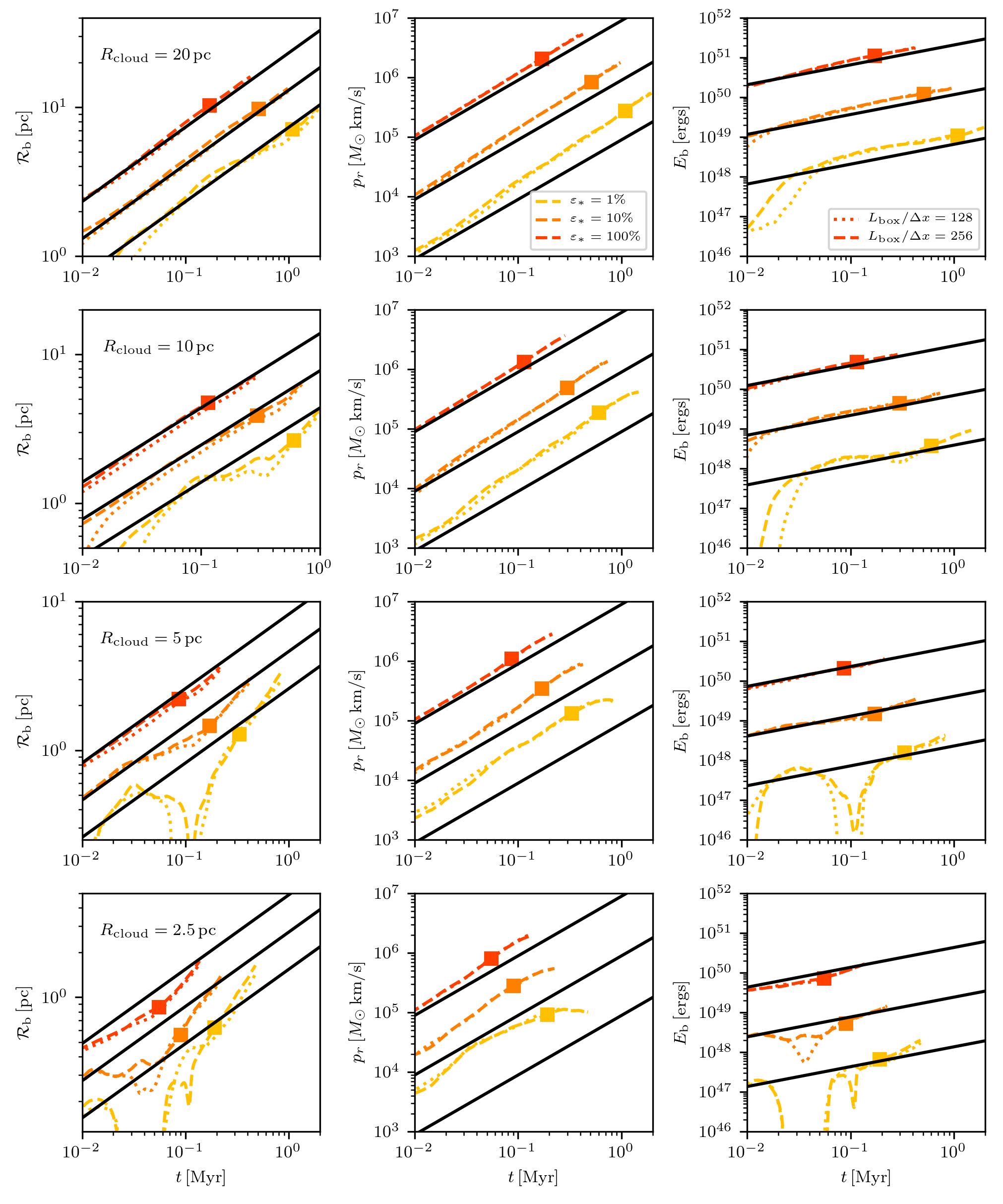}
    \caption{Same as \autoref{fig:m5e4_summary}  for cases with
    $\mcloud = 5\times 10^5 M_{\odot}$.}
    \label{fig:m5e5_summary}
\end{figure*}

\section{Results for Different Mass Clouds}
\label{app:different_mcloud}

In \autoref{tab:sim_params} we lay out the full range of 
simulations that we ran.  In the main body of the text we only 
displayed and discussed results for the cases with 
$\mcloud = 10^5 \, M_{\odot}$. In this appendix we display 
results from the runs with cloud masses of 
$\mcloud = 5\times 10^4 M_{\odot}$ and $5\times 10^5 M_{\odot}$, 
given in \autoref{fig:m5e4_summary} and \autoref{fig:m5e5_summary}, 
respectively. Results for both $128^3$ and $256^3$ resolution 
are shown.
%

\bibliography{bibliography}{}
\bibliographystyle{aasjournal}

\end{document}